\documentclass[iop,apj,tighten]{emulateapj}
\usepackage{apjfonts} 
\usepackage{amsmath,amstext}
\usepackage{graphicx}
\usepackage{rotating}
\usepackage{multirow}
\usepackage{threeparttable}
\usepackage[breaklinks,colorlinks,citecolor=blue,linkcolor=magenta]{hyperref} 
\usepackage[all]{hypcap} 

\begin{document}

\title{A large-scale search for evidence of quasi-periodic pulsations in solar flares}
\author{A. R. Inglis$^{1,2}$, J. Ireland$^{1,3}$, B. R. Dennis$^1$, L Hayes$^4$, P. Gallagher$^4$}
\affil{1. Solar Physics Laboratory, Heliophysics Science Division, NASA Goddard Space Flight Center, Greenbelt, MD, 20771}
\affil{2. Physics Department, The Catholic University of America, Washington, DC, 20064}
\affil{3. ADNET Systems Inc.}
\affil{4. Trinity College Dublin, Dublin, Ireland}

\begin{abstract}
The nature of quasi-periodic pulsations in solar flares is poorly constrained, and critically the general prevalence of such signals in solar flares is unknown. Therefore, we perform a large-scale search for evidence of signals consistent with quasi-periodic pulsations in solar flares, focusing on the 1 - 300s timescale. We analyse 675 M- and X-class flares observed by GOES in 1-8\AA\ soft X-rays between 2011 February 1 and 2015 December 31. Additionally, over the same era we analyse Fermi/GBM 15-25 keV X-ray data for each of these flares that was associated with a Fermi/GBM solar flare trigger, a total of 261 events. Using a model comparison method, we determine whether there is evidence for a substantial enhancement in the Fourier power spectrum that may be consistent with a QPP signature, based on three tested models; a power-law plus a constant, a broken power-law plus constant, and a power-law-plus-constant with an additional QPP signature component. From this, we determine that $\sim$ 30\% of GOES events and $\sim$ 8\% of Fermi/GBM events show strong signatures consistent with classical interpretations of QPP. For the remaining events either two or more tested models cannot be strongly distinguished from each other, or the events are well-described by single power-law or broken power-law Fourier power spectra. For both instruments, a preferred characteristic timescale of $\sim$ 5-30 s was found in the QPP-like events, with no dependence on flare magnitude in either GOES or GBM data. We also show that individual events in the sample show similar characteristic timescales in both GBM and GOES datasets. We discuss the implications of these results for our understanding of solar flares and possible QPP mechanisms.

\end{abstract}

\keywords{keywords}
\maketitle

\section{Introduction}
\label{introduction}

Over several decades, many studies have suggested the presence of characteristic timescales in the emission from solar flares. Collectively, these fluctuations are often referred to as quasi-periodic pulsations (QPP) or quasi-periodic oscillations (QPO). These fluctuations are typically observed on timescales of 1 s up to several minutes, and are seen over a wide range of wavelengths, from radio waves and microwaves \citep[e.g.][]{1970SoPh...13..420C, 2003ApJ...588.1163G, 2005A&A...439..727M, 2008A&A...487.1147I}, EUV \citep[e.g.][]{2012ApJ...749L..16D}, up to soft and hard X-ray \citep[e.g.][]{1969ApJ...155L.117P, 1983ApJ...271..376K, 2001ApJ...562L.103A, 2009SoPh..258...69Z, 2013ApJ...777...30I} and even gamma ray energies \citep[e.g.][]{2010ApJ...708L..47N}. Shorter \citep{2012ApJ...749...28T} and longer \citep{2010ApJ...719..151F} timescales are also present in flare emission, which are also sometimes referred to as QPP. Similar quasi-periodic signatures are also seen from stellar flares \citep[e.g.][]{2006A&A...456..323M, 2010ApJ...714L..98K, 2016MNRAS.tmp..660P} , where longer characteristic timescales are often observed \citep{2015MNRAS.450..956B, 2015ApJ...813L...5P}. However, despite many advances in instrumentation over the years, it remains unclear what physical mechanism is responsible for generating such characteristic timescales in flares. 

Two main mechanisms have gained traction as probable explanations for QPP \citep[see][for reviews]{2009SSRv..149..119N, 2016arXiv160902689V}. These are either that QPP are a signature of magnetohydrodynamic (MHD) wave modes \citep{1983SoPh...88..179E} generated in the flare arcade \citep[e.g.][]{2006A&A...452..343N, 2011ApJ...730L..27N}, or that they are a signature of the flare reconnection and energy release itself, which may occur in a bursty, or quasi-oscillatory manner \citep[e.g.][]{2006ApJ...642.1177L, 2006Natur.443..553D, 2009A&A...494..329M, 2016ApJ...820...60G}. Crucially however, most studies of this phenomenon have focused on a single event or a small sample of events, and there remains a lack of large-scale, statistically robust studies of QPP. As a result, regardless of the physical mechanism(s) responsible for these signals, their actual prevalence in flare emission is poorly known. 

Recent work has illustrated the ubiquity of power laws in the Fourier power spectra of a variety of astrophysical objects, including gamma-ray bursts \citep{2010AJ....140..224C}, magnetars \citep{2013ApJ...768...87H}, active galactic nuclei \citep{2006Natur.444..730M}, stellar and solar flares \citep{2011A&A...533A..61G, 2015ApJ...798..108I}, as well as emission from the Sun in coronal active regions and the quiet Sun \citep{2015ApJ...798....1I}. These power laws are believed to be an intrinsic property of the observed object, and it is therefore critical to take them into account when searching for significant oscillations or pulsations in the time series of such events. Failure to do so may lead to a large overestimation of the significance of power in the Fourier spectra of analysed signals \citep[see also][]{2005A&A...431..391V, 2010MNRAS.402..307V}. Additionally, the nature of the Fourier power spectrum in such signals means that the empirical subtraction of slowly varying components of the signal may lead to misleading results and should be avoided \citep{2010MNRAS.402..307V, 2011A&A...533A..61G, 2015ApJ...798..108I, 2016ApJ...825..110A}.

Considering this, \citet{2015SoPh..290.3625S} recently analysed 34 X-class flares from solar Cycle 24, while accounting for potential power-law properties of the Fourier spectrum. Their results suggest that oscillatory-like signals may indeed be prevalent in GOES (Geostationary Operational Environmental Satellite) soft X-ray data, at least during the impulsive phase of events, finding a signal in 25 out of 32 events. However, it is clear that further large-sample studies are needed.

In this work, we aim to address the question of the occurrence frequency of QPP signals in solar flares and their typical properties, focusing in particular on the 1 - 300 s range. We build on the previous efforts of \citet{2015ApJ...798..108I}, who described an analysis prescription that accounts for the power-law shape of the Fourier power spectrum. Here, we examine the lightcurves of all GOES M- and X- class flares occurring between 2011 February 1 and 2015 December 31, a sample of $\sim$ 700 events. For comparison, we find all of the events in this sample that were associated with a Fermi Gamma-ray Burst Monitor \citep[GBM,][]{2009ApJ...702..791M} event trigger. This subset comprises $\sim$ 300 events, which we analyse with GBM 15-25 keV X-ray data using the same methodology. To our knowledge this constitutes the first large-scale, self-consistent search for the presence of QPP in solar flares. Using this approach we can determine the occurrence frequency of such events, and the distribution of their estimated characteristic timescales.

\section{Instruments and data selection}
\label{methods}

\subsection{Instruments}
\label{instruments}

For this study we require a large sample of solar flare observations. Additionally, in order to be able to compare results across multiple wavelengths, we require a time period with continuous complementary observations. In our case, we have used data from the GOES instrument series, and from Fermi/GBM. For this reason, we choose the interval 2011 February 1 - 2015 December 31, as it not only coincides with the availability of GOES-15 satellite data, but also includes regular solar observations by GBM.

GOES satellites, in addition to their primary function as Earth-observing instruments, are equipped with solar X-ray detectors that record the incident flux in the 0.5 - 4 \AA\  and 1 - 8 \AA\ wavelength ranges. GOES observations are typically used to record the strength of a solar flare as a `GOES-class' on a logarithmic scale, which has become a standard flare descriptor in solar physics. Solar X-ray data from the most recent satellite, GOES-15, has been available since 2010 at a nominal 2s cadence. Due to the high geostationary orbit of GOES, data coverage is also almost continuous, which is ideal for this study.

Fermi is a gamma-ray and X-ray astrophysics mission launched in 2008 into low-Earth orbit. It has two scientific instruments; GBM, which operates in the 8 keV - 40 MeV range, and the Large Area Telescope \citep[LAT,][]{2009ApJ...697.1071A}, which observes in the 20 MeV - 300 GeV range. Both instruments have solar applications, but GBM in particular regularly observes emission from solar flares, with a solar duty cycle of $\sim$ 60\%, similar to the solar-dedicated Reuven Ramaty High Energy Solar Spectroscopic Imager \citep[RHESSI][]{2002SoPh..210....3L}. GBM consists of 12 NaI detectors and 2 BGO detectors, all at different orientations to ensure full-sky coverage and enable localization of emission sources \citep[see][]{2009ApJ...702..791M}. The NaI detectors provide sensitivity in the $\sim$ 8 keV - 1 MeV range, while the BGO detectors operate in the $\sim$ 250 keV - 40 MeV energy range. Similar energies are covered by RHESSI, which performs imaging spectroscopy between 3 keV - 17 MeV. However, RHESSI time series data is not ideal for this type of study. First, RHESSI rotates on its axis approximately every 4 s, as well as experiencing more complex second-order rotation effects that modify the time series data \citep{2011A&A...530A..47I}. Secondly, the aluminium attenuators used by RHESSI during periods of high count rates cause discontinuities in the time series data that are difficult to compensate for. Fermi/GBM time series data is relatively free from such artifacts. Accordingly, for this solar flare work, we focus on 15-25 keV X-ray data provided by the Fermi/GBM NaI detectors, as well as the GOES 1-8\AA\ data. 

For solar observations, it is necessary to select a particular GBM detector that is Sun-oriented during the flare. However, GBM can suffer from pile-up effects and spectral distortion during strong flares, because unlike RHESSI there are no attenuators available to reduce count rates during periods of strong emission. Given this, we construct a simple scheme using tools available in the SunPy data analysis package \citep{2015CS&D....8a4009S} to select the desired NaI detector for each event. First we determine the pointing angle between each detector and the Sun for the duration of the event. From these, we select the 5 detectors with the smallest angles to the Sun. As a second step, we select from these the detector with the lowest variance in angle to the Sun. Generally speaking, this results in the selection of a detector with an acceptable angle to the Sun, but not so close as to suffer from strong pile-up or spectral distortion effects. The low variance in the detector-Sun angle also ensures that long term systematic changes in the observed count rates from spacecraft motion are kept to a minimum. To ensure that the exact choice of detector does not have an undue influence on the results, we tested the use of different detectors on a selection of GBM events. We found that although there were slight variations in the determined parameters, the overall conclusions for each event were unaffected. The choice of detector is likely to  be more crucial at higher energies, e.g. above 50 keV, due to lower count rates.

\subsection{Data selection}
\label{data_selection}

To compile the GOES dataset, we choose all M- and X-class flares during the 2011 February 1 - 2015 December 31 time interval ($\sim$ 700 events), using the start and end times from the GOES catalogue as the time window for our analysis of each event. To access this catalogue, we use the Heliophysics Event Knowledgebase\footnote{\url{https://www.lmsal.com/hek/} } (HEK). As a preliminary step, we remove all short events from consideration, defined as an event with fewer than 200 data points, i.e. shorter than 400 s, as we do not consider our analysis method reliable for such short events. After this processing step, we are left with a database of 675 events.

For Fermi/GBM, as mentioned in Section \ref{instruments}, the low-Earth orbit of the spacecraft produces frequent occultations of the Sun, meaning necessarily fewer events are observed. To accumulate the database of Fermi/GBM events, we use the GBM trigger catalogue produced by the instrument team\footnote{See the Fermi Science Support Centre: \url{fermi.gsfc.nasa.gov/ssc/}}, selecting all events marked as flares. We cross-check this against the GOES catalogue, discarding any events where there is not a clear match between the two. Using this method, the total raw sample size for Fermi/GBM is 297 events (the actual number of events analysed is smaller, as discussed in Section \ref{gbm_results}). We choose the 15-25 keV range for study as most flares of M and X class emit significantly in X-rays in this energy range, whereas at higher energies fewer contain significant emission. 

For the purpose of our analysis, we use the CTIME data product produced by the instrument team, which provides high-cadence lightcurve data in pre-binned energy channels. For CTIME data, the nominal cadence is 0.256 s, with a burst mode cadence of $\approx$ 64 ms activated for several minutes during flare times. For this work, in order to achieve a uniform cadence and to ensure a reasonable signal-to-noise ratio, we rebin all data to 1 s cadence prior to analysis.

The onset of X-ray emission in higher-energy 15-25 keV Fermi data is generally more impulsive and of shorter duration than the lower-energy X-ray profile observed by GOES, due to the different energy ranges and characteristic exponential decay times of the observed emission. The GOES flux typically begins rising earlier than the emission in 15-25 keV X-rays, and also decays more slowly following the emission peak. Additionally, as the on-board GBM trigger is not designed specifically for flares, on many occasions the 15-25 keV X-ray flux during flares begins rising significantly in advance of the GBM trigger time. For these reasons, we require a different convention for setting the start and end times of the analysis for GBM. Hence, we construct the start and end times for GBM 15-25 keV analysis such that they are a sub-interval of the GOES analysis window, i.e. the GBM analysis window is fully contained within the GOES window. This allows us to directly compare the results from the two datasets (see Section \ref{compare_goes_gbm}). 

We select the start time as 120 s prior to the GBM burst trigger time. This is designed to ensure that any pre-trigger flare emission is captured in the majority of cases. Since the trigger catalogue does not contain event end times, we prescribe the end time to be midway between the GOES peak time and the GOES end time. This scheme is found in practice to well encapsulate the period of significant hard X-rays for flares. Additionally, this selection ensures that time window for GBM analysis is contained within the GOES time interval, although the GBM interval is usually shorter.

\section{Methodology}
\label{method}

The methodology for this work is based on that applied to solar flares in \citet{2015ApJ...798..108I}, and to solar active region data in \citet{2015ApJ...798....1I}. Here we describe the key points of the method, including modifications from that work. The first step in the analysis procedure is to normalize the input data, such that,

\begin{equation}
I_{norm} = \frac{I - \bar{I}}{\bar{I}}
\end{equation}

where $I$ is the original signal and $\bar{I}$ is the mean of the signal. This operation normalizes the signal only - no subtraction of an estimated smoothed background component or time-differencing is performed, unlike in other works \citep[e.g.][]{2012ApJ...749L..16D, 2015SoPh..290.3625S, 2016Dennis}. This is important as it ensures that no artifacts are introduced to the Fourier power spectrum of the signal, which could yield misleading results \citep[e.g.][]{2010MNRAS.402..307V, 2011A&A...533A..61G, 2016ApJ...825..110A}. The next step is to apodize this normalized signal by choosing an appropriate window function, to mitigate the effects of the finite-duration time series on the Fourier power spectrum. In this work, we utilize the Hann window \citep{Blackman:1959:MPS} function. Other functions, such as the Blackman-Harris window or the Hamming window would be equally appropriate. In general, the results are not very sensitive to the choice of window function \citep[see][]{2015ApJ...798..108I}. Apodizing has the side-effect of broadening peaks in the Fourier power spectrum, meaning that neighbouring frequencies may no longer be independent. However, given the typical number of frequencies in flare Fourier power spectra, and that the features we analyse typically encompass at least several frequencies (see Section \ref{results}), this should not impact the results. 

After these data preparation steps, the main element of the procedure is to perform a model comparison on the Fourier power spectrum of each solar flare. In this work we consider three models; a single power-law plus constant model ($S_0$), a power-law-plus-constant model with an additional localized enhancement ($S_1$), and a broken-power law model ($S_2$). These models may be written:

\begin{equation}
S_0(f) = A_0 f^{-\alpha_0} + C_0
\label{m0_eqn}
\end{equation}

\begin{equation}
S_1(f) = A_1 f^{-\alpha_1} + B \ \exp \left( \frac{-(\ln f - \ln f_p)^2}{2\sigma^2} \right) + C_1
\label{m1_eqn}
\end{equation}

\begin{equation}
  S_2(f)=\begin{cases}
    A_2 f^{-\alpha_{b}} + C_2, & \text{if $f<f_{break}$}.\\
    A_2 f^{-\alpha_{b} - \alpha_{a}} f^{-\alpha_{a}} + C_2, & \text{if $f>f_{break}$}.
  \end{cases}
  \label{m2_eqn}
\end{equation}

The choice of model $S_0$ is based on the observation that power-law Fourier power spectra are a common property of a variety of astrophysical and solar objects \citep{2010AJ....140..224C, 2011A&A...533A..61G, 2013ApJ...768...87H, 2015ApJ...798..108I, 2015ApJ...798....1I}, and that such power laws can lead naturally to the appearance of bursty features in time series \citep[e.g.][]{2015ApJ...798....1I}. This power law must be accounted for in the model in order to avoid a drastic overestimation of the significance of peaks in the power spectrum \citep{2005A&A...431..391V, 2011A&A...533A..61G}. The constants $C_0$, $C_1$ and $C_2$ account for a transition between power law behaviour and `white noise' behaviour in the Fourier power spectrum. The second model $S_1$ is equivalent to model $S_0$ plus an extra term corresponding to a Gaussian enhancement in log-frequency space. This model component is designed to represent a signature that may be consistent with classical ideas of QPP, i.e. excess power in a localized frequency range. Note that this model is empirical, but is derived from our expectation that due to the rapid evolution of solar flares in the impulsive phase, any QPP signature would necessarily have some width in Fourier space. In studies of the solar interior, for example identification of solar p-modes, theoretically based models have been tested \citep[e.g.][]{1990ApJ...364..699A, 2012MNRAS.419.1197K}. However, solar flares are much more impulsive and rapidly evolving phenomena, and an appropriate model choice is unclear. Model $S_2$ is motivated by the idea that a single power-law model (i.e. $S_0$) may be too restrictive to fit flare Fourier power spectra, and that the presence of an enhancement in the spectrum is not necessarily best described by model $S_1$. The potential for a spectral break allows broad features in the Fourier spectrum to be captured without resorting to a `QPP-like' model. 

For model $S_1$, we impose the following constraints on the values of the parameters, in particular $f_p$ and $\sigma$, in order to restrict our range of interest. For this reason, we choose:

\begin{equation}
\begin{split}
1 < P < 300 \textnormal{ s} , \\
0.05 < \sigma < 0.25
\end{split}
\label{limits}
\end{equation}

where $\sigma$ is the width of the peak in log-frequency space, and $P$ = 1/$f_p$, i.e. the model only considers characteristic timescales in the 1-300s range. Shorter periods are not detectable by GOES or Fermi/GBM due to the instrument time resolution. while longer periods we consider out of the scope of this analysis. Additionally, the 2 s time cadence of GOES data means that in reality only periods of at least 4 s can be detected in the Fourier domain for this instrument. Similarly, with our GBM data binned to 1 s the minimum detectable period is 2 s. The limitations on the width parameter are necessary to ensure that the enhancement in Fourier power covers multiple datapoints, but is constrained to be localized. Without such a constraint, the bump component may take on an arbitrarily large width and dominate the entire spectral window.

In order to find the best fit to each model, we use model fitting tools provided by the \verb|SciPy| data analysis package. From this we determine the maximum likelihood $L$ for each model and the associated best-fit parameters. The likelihood function may be written \citep[e.g.][]{2005A&A...431..391V}:

\begin{equation}
L = \prod^{N/2}_{j=1} \frac{1}{s_j} \exp \left(-\frac{i_j}{s_j} \right)
\end{equation}

where $I$ = ($i_1$,...,$i_{N/2}$) represents the observed Fourier power at frequencies $f_j$ for a time series of length $N$, and $S$ = ($s_i$,...,$s_{N/2}$) represents the model of the Fourier power spectrum, either $S_0$, $S_1$ or $S_2$. In order to ensure that we adequately cover the available parameter space, each fit is repeated 20 times with randomized initial guess values, and the parameters that yield the largest value of $L$ are retained as the final best fit. This procedure helps to avoid the problem of fitting local maxima in $L$, but not finding the true global maximum likelihood.

In the next analysis step, in order to compare which model is more appropriate for the observed data, we use the Bayesian Information Criterion (BIC). This criterion \citep{Burnham01112004} is defined by:

\begin{equation}
BIC = - 2 \ln(L) + k \ln(n)
\label{bic_eqn}
\end{equation}

where $L$ is the maximum likelihood, $k$ is the number of free parameters in the model and $n = N/2$ is the number of data points in the Fourier power spectrum. The key concept of BIC is that there is a built-in penalty for adding complexity to the model. This is adopted because a model with more parameters should always better maximize the likelihood - or equivalently minimize the negative log-likelihood - than a model with fewer parameters in the case where the simpler model is a subset of the more complex one, as is the case here. Using the BIC value to compare models therefore tests whether added complexity is sufficiently justified. 

A smaller value of BIC indicates that a model is preferred over others. To compare models therefore, we calculate $\Delta$BIC between the two models. In general, a $\Delta$BIC value >10 is considered strong evidence in favour of one model over another \citep{Burnham01112004}. In this work we use the following criterion to determine whether a model $i$ is `strongly favoured':

\begin{equation}
BIC_i < BIC_j - 10
\label{strong_test}
\end{equation}

for all other models $j \ne i$.

It is important to note however that neither likelihood or BIC determine whether a model is actually a good choice in absolute terms - all of the tested models could be unrepresentative of the observed data. In order to determine if a model is actually consistent with the data we require a goodness-of-fit statistic. In this work, we utilize the $\chi^2$-like statistic for exponentially distributed data as described by \citet{2014ApJ...789..152N}, who showed that an appropriate statistic may be written:

\begin{equation}
\chi^2_{\nu} = \frac{1}{\nu} \sum^{n}_{j=1} (1 - \hat{\rho_j})^2 
\label{chi2_like}
\end{equation}

for a single Fourier power spectral sample, where $\hat{\rho_j} = i_j / {s_j}$ is the sample-to-model estimator, i.e. the ratio of the sample data to the best-fit model \citep[see][Equation 5]{2014ApJ...789..152N}, and $\nu$ is the number of degrees of freedom. For each model, we determine $\chi^2_{\nu}$ and using the approximate expression for the probability density function of $\chi^2_{\nu}$ \citep[][Appendix A]{2014ApJ...789..152N} subsequently find the probability $p$ of the data given the chosen model. For these events where no model provides an adequate fit (defined in this work as $p < 0.01$), the model comparison via $\Delta$BIC has limited value; in these cases we can say only that additional models should be tested. This approach differs from that adopted in \citet{2015ApJ...798..108I}, where an MCMC method was used to numerically generate the underlying distributions of chosen test statistics \citep[see also][]{2010MNRAS.402..307V}. The MCMC method, while ideal, is too numerically intensive to be viable for this large sample of events.

\section{Results}
\label{results}

\subsection{GOES soft X-ray data}
\label{goes_results}

For each of the 675 events in our GOES sample, we examined the emission from the GOES 1-8\AA\ channel, fit models $S_0$, $S_1$ and $S_2$ to the Fourier power spectral density, and performed the model comparison test described in Section \ref{method}. Therefore, for each event we determine the values of BIC and $\chi^2_{\nu}$ for each model, using $\Delta$BIC to establish which of the tested models was preferred, and $\chi^2_{\nu}$ to determine how well both models fit the data. 

\begin{figure*}
\begin{center}
\includegraphics[width=18cm]{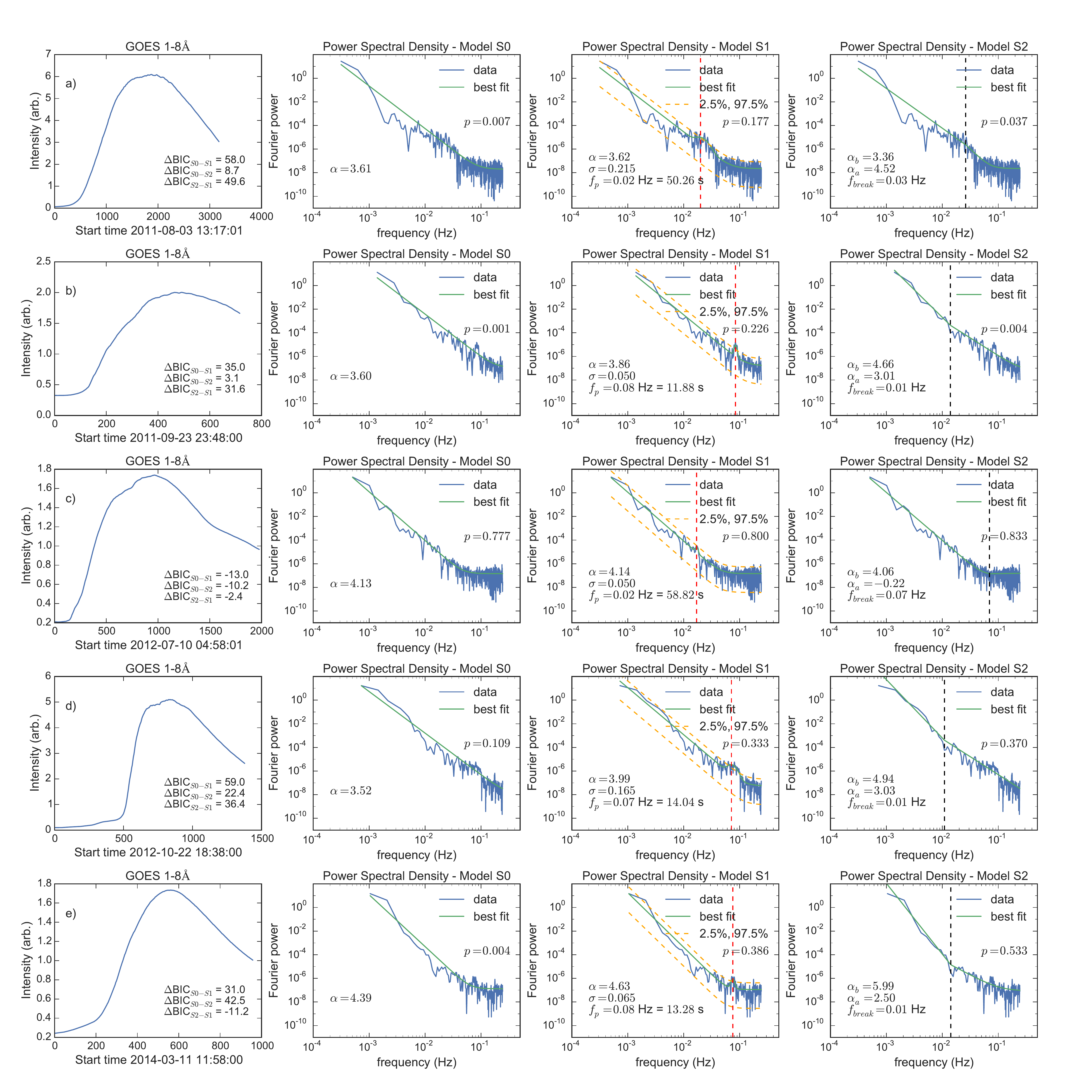}
\caption{Examples of the model comparison analysis applied to a selection of flares observed by GOES in the 1-8\AA\ range on a) 2011-08-03, b) 2011-09-23, c) 2012-07-10, d) 2012-10-22, and e) 2014-03-11. For each event, the original time series signal is shown in the left panel, while the fits to each model $S_0$, $S_1$ and $S_2$ are shown in the remaining panels. The vertical dashed red line in the third panels denotes $f_p$, the location of the additional spectral bump. The black vertical line in the fourth panels shows the location of the break frequency $f_{break}$. The p-values corresponding to the $\chi^2$-like values obtained via Equation \ref{chi2_like} are shown in the upper right of each panel. As the $\Delta$BIC values show, flares a,b and d are categorised as strongly favouring model $S_1$, while flare c) favours the single power law model $S_0$ and flare e) strongly favours the broken power law model $S_2$.}
\label{goes_examples}
\end{center}
\end{figure*}

Typical examples of the model comparison analysis are shown in Figure \ref{goes_examples}. The original input data are shown in the left panels for each event, while the fits of models $S_0$, $S_1$ and $S_2$ to the Fourier power spectra are shown in the remaining panels. For model $S_1$, the 2.5\% and 97.5\% quantiles relative to the power-law component of the model are shown, illustrating the height of the additional bump component. The peak location of the bump is denoted by the vertical red line. In general, when there is a lack of evidence for model $S_1$, the best-fit amplitude $B$ of this extra component often goes to zero (its minimum allowed value), resulting in almost identical fits for models $S_0$ and $S_1$. By contrast, when model $S_1$ is preferred, there is usually a substantial additional peak visible in the spectrum which is several $\sigma$ above the mean of the power-law component.

Based on the $\Delta$BIC test, We separate all the surveyed events into four categories; S0, S1, S2, and U, or unresolved. Each category is populated by events where that model is strongly preferred over all others according to Equation \ref{strong_test}. For example, an event is classified as S0 if model S0 produces the lowest value of BIC, and the $\Delta$BIC relative to both other models is at least 10 in favour of $S_0$. Additionally, to confirm an event in the S0 category the goodness-of-fit $\chi^2_{\nu}$ must be satisfactory, defined as having a $p$-value > 0.01. If the $p$-value for the given model fails this test, it is excluded from the category.  The same criteria apply for categories S1 and S2. The remaining events are classified as unresolved events - these are cases where at least two of the tested models cannot be strongly distinguished from each other.

\begin{figure*}
\begin{center}
\includegraphics[width=18cm]{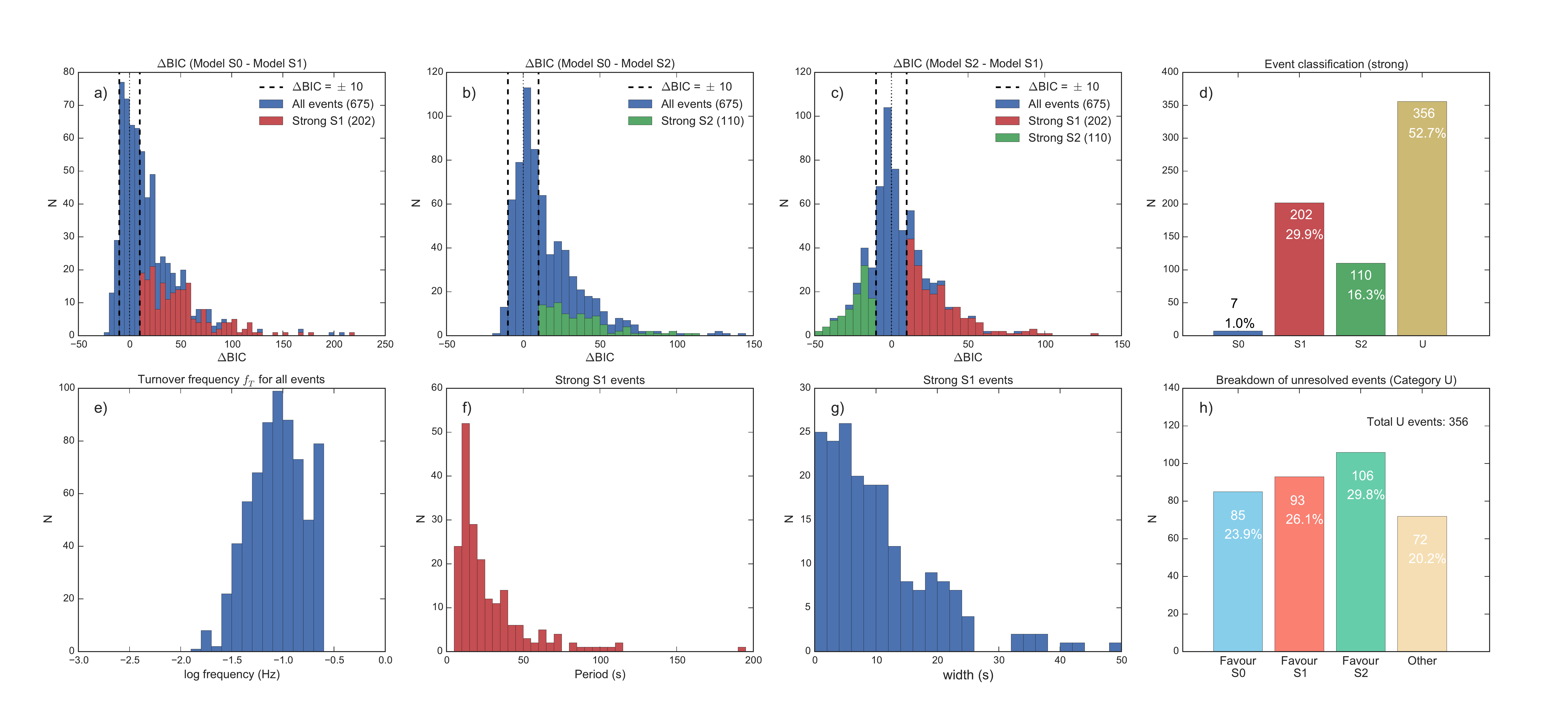}
\caption{Distributions of various parameters from the model comparison analysis of 675 GOES events. \textit{a)} Distribution of $\Delta$BIC values for the comparison between model $S_0$ and model $S_1$ for each flare. The vertical dashed lines denote $\Delta$BIC $\pm$ 10. \textit{b)} Distribution of $\Delta$BIC values for the comparison between model $S_0$ and model $S_2$ for each flare. \textit{c)} Distribution of $\Delta$BIC values for the comparison between model $S_2$ and model $S_1$ for each flare. In each of these three panels, events that meet all the criteria needed to assign a strongly preferred model are highlighted, with S1 events shown in red and S2 events shown in green. \textit{d)} Breakdown of the event classifications according to their $\Delta$BIC values. Events where model $S_0$ is strongly preferred over all other models are labeled `S0'. Events where model $S_1$ is strongly preferred are labelled `S1', and where model $S_2$ is preferred the event is categorised as `S2'. Events where at least two models may be appropriate are classified as unresolved, or U. \textit{e)} Distribution of Fourier power spectrum turnover frequency $f_T$ for all events. \textit{f)} Distribution of characteristic timescales - or periods - for all S1 events.
\textit{g)} Distribution of characteristic width for all S1 events, converted into seconds using $\pm \sigma$ values. \textit{h)} Breakdown of the event classification when the model selection criterion is relaxed to $\Delta$BIC > 0, i.e. the distribution of which model is favoured, regardless of margin.}
\label{afino_distributions}
\end{center}
\end{figure*}

The full survey results are summarized in the distributions shown in Figure \ref{afino_distributions}, with the main findings also shown in Table \ref{summary_table}. Figures \ref{afino_distributions}a, \ref{afino_distributions}b and \ref{afino_distributions}c show the histograms of $\Delta$BIC values for all three model comparisons; $S_0$ - $S_1$ (panel \ref{afino_distributions}a), $S_0$ - $S_2$ (panel \ref{afino_distributions}b), and $S_2$ - $S_1$ (panel \ref{afino_distributions}c), for all 675 events. Figure \ref{afino_distributions}d shows the categorisation of these events as described above; we find 7 S0 events (1\%), 202 S1 events (30\%), 110 S2 events (16\%), with the remaining 356 (53\%) events unresolved. These unresolved events are further examined in Figure \ref{afino_distributions}h. Hence, of the 675 events surveyed, 202 show evidence of an additional component in the Fourier power spectrum, that may be consistent with a QPP signature. We can see that the distributions of $\Delta$BIC are not symmetric about the peaks. For example, in the comparison between model $S_0$ and $S_1$ shown in Figure \ref{afino_distributions}a, the mode of the distribution is $\Delta$BIC $\sim$ -8, but there is a substantial tail of events showing positive $\Delta$BIC values, with a few events showing values $>$ 200. By contrast the values of $\Delta$BIC cut-off more sharply on the negative side of the distribution, with the smallest recorded value of $\Delta$BIC $\approx$ -21. This cut-off can be naturally explained by considering that model $S_1$ reduces to $S_0$ when $B \to 0$, hence the likelihood $L$ for $S_1$ should always be at least that of $S_0$. Hence for cases where no enhancement is found in the Fourier power spectrum, the likelihood of the best fit models $S_0$ and $S_1$ will be equal, with the difference in $BIC$ being determined by the term $k \ln(n)$ (see Equation \ref{bic_eqn}).

\begin{figure*}
\begin{center}
\includegraphics[width=18cm]{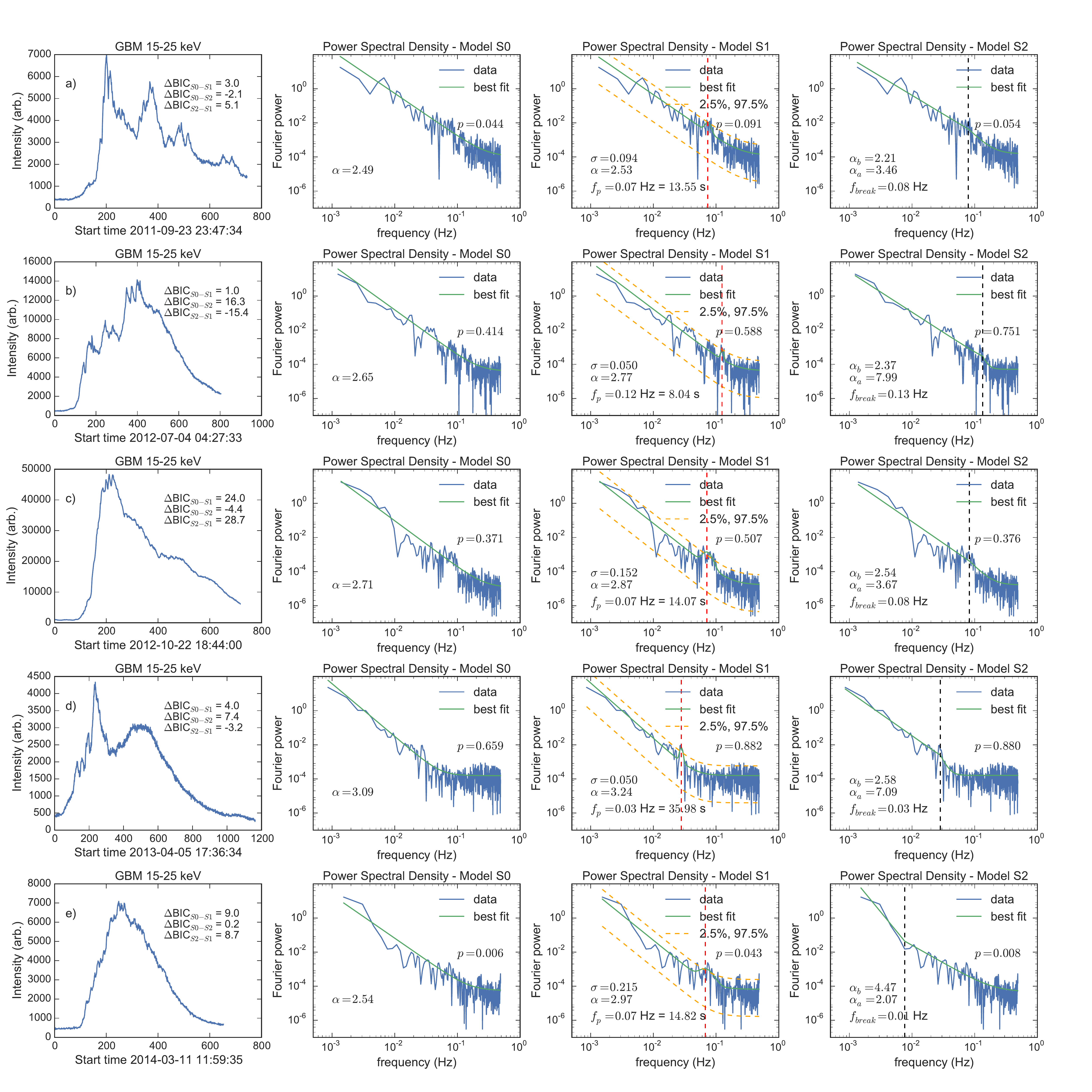}
\caption{Examples of the model comparison analysis applied to a selection of flares observed by Fermi/GBM in the 15-25 keV range on a) 2011-09-23, b) 2012-07-04, c) 2012-10-22, d) 2013-04-05, and e) 2014-03-11. For each event, the original X-ray time series signal is shown in the left panel, while the fits to each model $S_0$, $S_1$ and $S_2$ are shown in the remaining panels. The vertical dashed red line in the third panels denotes $f_p$, the location of the additional spectral bump. The black vertical line in the fourth panels shows the location of the break frequency $f_{break}$. Flares a), d), and e) find that at least two models are competitive, while Flare b) strongly favours model $S_2$ and flare c) strongly favours model $S_1$.}
\label{gbm_examples}
\end{center}
\end{figure*}

Figure \ref{afino_distributions}f shows the distribution of characteristic timescales, or `periods', for those events where model $S_1$ was strongly preferred over both the single power-law model $S_0$ and the broken power law model $S_2$. We can see that a majority of the survey events show a best-fit period in the 10-30 s region, with the median value of the distribution $P_{median}$ = 17.5 s and the modal value $P_{mode}$ $\approx$ 12 s. An important caveat is that there may be a selection effect at short values of $P$ that cause fewer events to be observed. This is due to the 2s time cadence of GOES data. This means that events with $P$ < 4 s cannot be observed in this analysis, while there may be a selection bias against events with $P \lesssim 10$. Hence we cannot say whether the period distribution for GOES truly peaks at $\sim$ 12 s. 

This can be illustrated by examining the turnover frequency $f_T$ (see Figure \ref{afino_distributions}e), the point at which the best fit model Fourier power spectrum transitions from a power-law slope to a constant background (see Equations \ref{m0_eqn}, \ref{m1_eqn}, \ref{m2_eqn}). Here we define $f_T$ as the point where the amplitude of the best-fit power law becomes equal to the best-fit constant $C$, i.e. for model $S_0$ the point at which $A_0f^{-\alpha_0}$ = $C_0$. The location of $f_T$ shows where the underlying noise level begins to dominate the flare signal, affecting how likely a periodic signal is to be detected. The distribution is peaked around a modal value of $f_T \sim$ 0.1 Hz, corresponding to $P \sim$ 10 s. Hence in many events the data limits the possibility of detecting a $P$ < 10 s signal. Figure \ref{afino_distributions}e also shows that there is a substantial edge of events with $f_T$ = 0.25 Hz, corresponding to the shortest frequency in the GOES signal. For these events, there is no discernible transition to a white noise regime.

At longer periods, few events are detected with characteristic timescales in the 100 - 300s range, where 300 s is the upper limit on $P$ that we have defined in our parameter search. These results indicate a strong preference for characteristic signals in the 10-30s range in GOES data. This distribution of periods is also consistent with a pattern that has been previously speculated to be present in GOES data \citep{2015SoPh..290.3625S, 2016Dennis, 2016Hayes}.

In Figure \ref{afino_distributions}g we examine how the periods and peak widths of the S1 events are distributed. Here we represent the peak widths using as the $\pm$ 1$\sigma$ values of $f_p$ translated into period space. We can see that although many events indicate relatively localized structures, characteristic widths of 15-20 s are common, with a small number of events showing widths $>$ 30 s. One possibility is that these broadband events are evidence of rapidly evolving timescales in flares, or an indication of the presence of multiple co-temporal signatures. However, we have ruled out the possibility that a larger structure such as a spectral break in the Fourier power spectral density is being incorrectly fit by $S_1$; such events would show a preference for the broken power-law model $S_2$.

As Figure \ref{afino_distributions}d shows, using our strict classification criteria over 50\% of studied events fall in the `unresolved' category. In Figure \ref{afino_distributions}h we examine these further by relaxing the model selection criteria simply to $\Delta$BIC > 0. Hence, for each of the unresolved events, we can see which of the three models is favoured. Of the 356 unresolved events, we find that 85 (24\%) favour model $S_0$, 93 (26\%) favour model $S_1$, 106 (30\%) favour model $S_2$, with the remaining 72 (20\%) flares unassigned. These unassigned events occur when the most favoured model fails the goodness-of-fit criterion described in Section \ref{method}. If we combine the strongly classified events in Figure \ref{afino_distributions}d with these additional events, we find that 92 events favour or strongly favour model $S_0$, with 295 for model $S_1$, and 216 for model $S_2$. This information is summarised in Table \ref{summary_table}. 

For this work, we prefer the strict $\Delta$BIC > 10 criterion shown in Equation \ref{strong_test} for the purposes of identifying potential QPP events. Therefore, we can say that at approximately 30\% of studied events show a signal in GOES consistent with the presence of QPP, and that there appears to be a clear preferential timescale associated with signals in GOES flare data. We examine the properties of events that `favour' model $S_1$ further in Sections \ref{compare_goes_gbm}, \ref{data_exploration}.

\subsection{Fermi/GBM X-ray data}
\label{gbm_results}

It is highly desirable to compare the GOES 1-8 \AA\ survey results with those obtained at other wavelengths, since QPP are known to be detected over a wide range of wavelengths (see Section \ref{introduction}). Studying QPP in different energy ranges can yield insight into the physical mechanism responsible for their generation. For this reason, we apply the same analysis method to Fermi/GBM 15-25 keV data for all observed events between 2011 February 1 and 2015 December 2015, as described in Section \ref{data_selection}. 

In addition to the caveats described in Section \ref{data_selection}, we find that, some analysed events show discontinuities in the lightcurves, either due to missing data, occultation of the Sun, or an observation mode change of the instrument. These discontinuities cause large distortions in the Fourier power spectrum, invalidating the analysis method. Using a manual search, we find 36 of these events, which we discard from consideration. This leaves us with a final Fermi/GBM data set of 261 analysed events for consideration. Select examples are shown in Figure \ref{gbm_examples}.

\begin{figure*}
\begin{center}
\includegraphics[width=18cm]{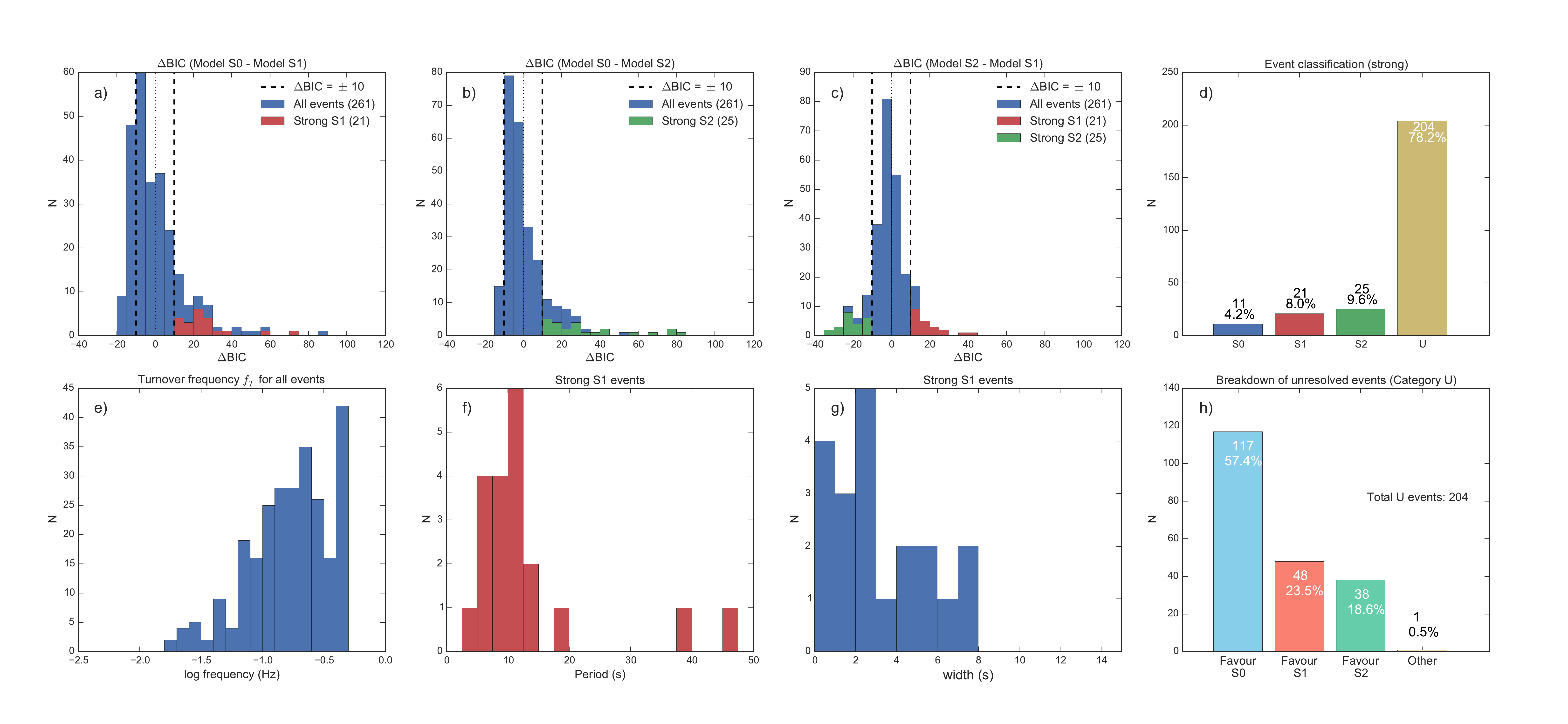}
\caption{Distributions of various parameters from Fermi/GBM 15-25 keV X-ray data survey results, totalling 261 events. \textit{a)} Distribution of $\Delta$BIC values for the comparison between model $S_0$ and model $S_1$ for each flare. The vertical dashed lines denote $\Delta$BIC $\pm$ 10. \textit{b)} Distribution of $\Delta$BIC values for the comparison between model $S_0$ and model $S_2$ for each flare. \textit{c)} Distribution of $\Delta$BIC values for the comparison between model $S_2$ and model $S_1$ for each flare. In each of these three panels, events that meet all the criteria needed to assign a strongly preferred model are highlighted, with S1 events shown in red and S2 events shown in green. \textit{d)} Breakdown of the event classifications according to their $\Delta$BIC values, using the same criteria as GOES (Figure \ref{afino_distributions}). \textit{e)} Distribution of Fourier power spectrum turnover frequency $f_T$ for all events. \textit{f)} Distribution of characteristic timescales - or periods - for all S1 events.\textit{g)} Distribution of characteristic widths in seconds for all S1 events. \textit{h)} Breakdown of the event classification when the model selection criterion is relaxed to $\Delta$BIC > 0, i.e. the distribution of which model is favoured, regardless of margin.}
\label{afino_distributions_gbm}
\end{center}
\end{figure*}

The resulting distributions for this dataset are shown in Figure \ref{afino_distributions_gbm}, with the main results summarized in Table \ref{summary_table}. As before, Figures \ref{afino_distributions_gbm}a, b, c show the $\Delta$BIC distributions for each of the three axes of model comparison. As with the GOES results, events which strongly favour model $S_1$ are highlighted in red, while events that strongly favour $S_2$ are highlighted in green. Figure \ref{afino_distributions_gbm}d shows the breakdown of classifications for all events, based on the strict criterion of Equation \ref{strong_test}. The difference between these GBM results and their GOES counterparts is substantial - most obvious is that a much larger fraction of events fall in the `unresolved' category, over 78\% of the sample. Accordingly, we find only 21 events categorised as S1 (8\%), 25 as S2 (10\%) and a further 11 as S0 (4\%). 

Inspection of the distribution of periods for these S1 events (as shown in Figure \ref{afino_distributions_gbm}f) shows a similar distribution to that seen in GOES however. For the GBM dataset, we find $P_{median}$ = 10 s and $P_{mode} \approx$ 11 s, shorter than the 17.5 s and 12 s values found in GOES. This may be due to the finer time resolution of the GBM data, with $dt$ = 1 s, which enables events with $P$ < 4 s to be detected, unlike in GOES data. At longer periods, only one event is found with $P>40$s. As before, Figure \ref{afino_distributions_gbm}e shows the distribution of turnover frequency $f_T$ for the GBM events. Similar to GOES, we see a substantial edge of events at the highest available frequency, indicating that for these events a transition to a white-noise regime never occurs. For the remaining events however, the modal value of the distribution is at $f_T \sim$ 0.2 Hz, a substantially higher frequency than was found in the GOES dataset. Hence it may be easier to detect higher frequencies in the GBM data compared with GOES.

For the GBM data, it is particularly interesting to understand the properties of the events that were classified as `unresolved', as these account for over three-quarters of the studied events. In Figure \ref{afino_distributions_gbm}h we show which model was favoured in these cases. In contrast with the GOES results, here the single power-law model $S_0$ is much more likely to be favoured, accounting for 117 of the unresolved events, or 57\%. 48 events favour model $S_1$ (24\%), and 38 favour model $S_2$ (19\%). A similar pattern holds if these events are combined with the strongly classified events in Figure \ref{afino_distributions_gbm}d; we see that in total 128 out of 261 events favour or strongly favour $S_0$, with 69 for model $S_1$ and 63 for $S_2$.

In general, the GBM data supports the simplest model $S_0$ substantially more often than the GOES dataset, and correspondingly finds a smaller fraction of events showing evidence for QPP signatures. The 21 S1 events comprise only $\sim$ 8\% of the total sample of 261 available in the Fermi/GBM data, compared with almost 30\% in the GOES sample. Figure \ref{afino_distributions_gbm}f shows that the detected characteristic timescales are distributed similarly to those observed in GOES, with a strong preference for a signal in the 5-15 s range, although their characteristic widths tend to be narrower. Several factors may explain the differences between the GOES and GBM surveys. The lower detection rates in GBM may be a result of the lower signal-to-noise ratio in the 15-25 keV X-ray data relative to GOES soft X-rays. GBM 15-25 keV X-rays are also sourced from different flare plasma and include a non-thermal component, compared to the soft X-rays observed by GOES, which observe only the thermal emission from hot coronal plasma.  

\begin{table}
\centering
\caption{Summary of GOES and GBM sample statistics}
\label{summary_table}
\begin{tabular}{c|r|r|r}
                          &                                  & GOES           & GBM         \\
\hline
\hline
  &  Total number of events:              & 675           & 261                          \\
\hline
  
\multirow{3}{*}{S0}   	  & \textbf{Strongly Favoured}  			& \textbf{7   (1.0\%)  }           & \textbf{11 (4.2\%) }    \\
                          & Favoured (any margin)       & 92   (13.6\%)         & 128 (49\%)       \\
                          & $\bar{\alpha}_0$             & 3.95 $\pm$ 0.41       & 2.95 $\pm$ 0.3  \\
                          \hline
\multirow{5}{*}{S1}       & \textbf{Strongly Favoured}           & \textbf{202 (30\%)  }          & \textbf{21 (8.0\%) }\\
                          & Favoured (any margin)       & 295 (43.7\%)          & 69 (26\%) \\
                          & $\bar{\alpha}_1$             & 4.28 $\pm$ 0.25       & 3.12 $\pm$ 0.38 \\
                          & $P_{median}$				& 17.5 s				& 11 s \\
                          & $P_{mode}$					& 12 s					& 8 s \\
                          \hline
\multirow{4}{*}{S2}       & \textbf{Strongly Favoured}           & \textbf{110 (16.3\%)}          & \textbf{25 (9.6\%)} \\
	                      & Favoured (any margin)       & 216 (32\%)            & 63 (24\%) \\
                          & $\bar{\alpha}_{b}$          & 5.39 $\pm$ 0.51       & -            \\
                          & $\bar{\alpha}_{a}$          & 2.82 $\pm$ 0.27       & -

\end{tabular}
\begin{tablenotes}
 \item `Strongly favoured' refers to events that prefer a model $i$ by a margin of BIC$_i$ < BIC$_j$ - 10, for $j \ne i$ (see Equation \ref{strong_test}). 
 \item Favoured by any margin refers to events where BIC$_i$ < BIC$_j$ for $j \ne i$. 
\end{tablenotes}

\end{table}

\subsection{Comparison between GOES and Fermi/GBM results}
\label{compare_goes_gbm}

An important question to address is whether GOES and Fermi/GBM, observing different energy ranges and hence different flare plasma, yield similar results. Although we have seen from Section \ref{goes_results} and \ref{gbm_results} that the overall distributions of $\Delta$BIC and characteristic timescale $P$ are similar for both instruments, it is possible that the results for individual events may be quite different. GOES observes only thermal plasma at lower temperatures, while Fermi/GBM is generally observing higher temperature plasma, as well as non-thermal emission that may be associated with flare footpoints. 

\begin{figure*}[ht]
\begin{center}
\includegraphics[width=18cm]{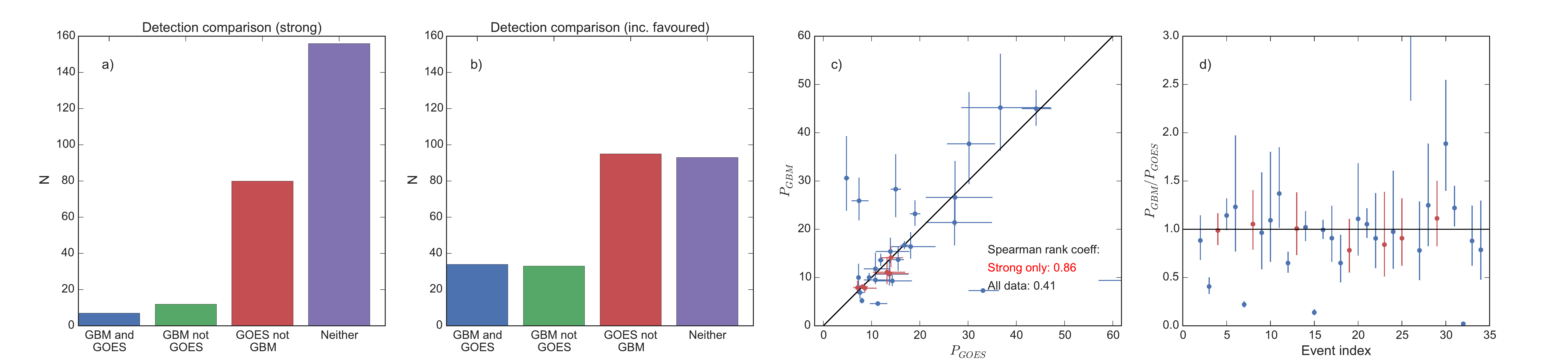}
\caption{Comparison of measured parameters in Fermi/GBM and GOES data. a) All 261 analysed events co-observed by Fermi/GBM and GOES split into four categories, 1) A QPP-like signature (strong preference for model $S_1$) in both GBM and GOES, 2) a strong signature in GBM but not GOES, 3) a strong signature in GOES but not GBM, and 4) no strong signature in either instrument. b) Same as panel a) but with the model selection criterion relaxed to $\Delta$BIC > 0. Hence the four categories are 1) tentative evidence for a signal in GOES and GBM, 2) tentative evidence for a signature in GBM but not GOES, c) tentative evidence for a signature in GOES but not GBM, and 4) No evidence in either instrument. c) $P_{GBM}$ vs $P_{GOES}$ for the 34 events showing some preference for model $S_1$ in both instruments. The 7 events showing strong evidence for model $S_1$ in both instruments are marked in red. The line $y = x$ is shown for reference. Here the error bars represent the best-fit width $\sigma$ for the event in each instrument. The Spearman rank correlation coefficient is shown both for all 34 tentative events (0.41) and for the 7 strong events (0.86). d) The ratio $P_{GBM}$ / $P_{GOES}$ for the same events, with the 7 strong events marked in red, as in the third panel. }
\label{gbm_vs_goes}
\end{center}
\end{figure*}

To investigate the relationship between the GOES and Fermi/GBM results, we compare the detection of QPP-like signatures in both instruments, i.e. whether an event classed as S1 in the Fermi/GBM data set is also classified as S1 in the GOES dataset. The results are shown in Figure \ref{gbm_vs_goes}. We can see that, of the 261 events that were co-observed by both instruments, 7 were classed as S1 by both GBM and GOES, while a further 14 were classed as S1 in GBM but not GOES. Conversely, 80 events were recorded as S1 in GOES but not GBM, and the remaining 156 events were not classed as S1 for either instrument.

It is also of interest to examine the additional events that favour model $S_1$ for both instruments, even though these do not meet our strict criteria of BIC$_1$ < BIC$_{j}$ - 10 for $j$ = 0,2. Figure \ref{gbm_vs_goes}b shows that when the model selection criteria are relaxed to $\Delta$BIC > 0, as we showed in Figures \ref{afino_distributions}, \ref{afino_distributions_gbm}, we find that there are 34 events that prefer model $S_1$ in both GOES and GBM by any margin, while 33 prefer $S_1$ in GBM but not GOES, 95 events prefer model $S_1$ in GOES but not GBM, and the remaining 93 prefer an alternative model in both instruments. 

Hence, events where a QPP signature is suggested in the Fermi/GBM dataset do not necessarily have corresponding signatures in GOES. Similarly, in reverse it is also common that a GOES signature will not have a clear counterpart in the GBM dataset. In general, Figures \ref{afino_distributions} and \ref{afino_distributions_gbm} suggest that overall there is greater sensitivity to such QPP-like signals in GOES data, compared to the higher energy 15-25 keV data obtained by Fermi/GBM. This is despite indications (see Figure \ref{afino_distributions}e) that short-period signals (< 10 s) may be easier to detect with GBM.

Panel \ref{gbm_vs_goes}c shows the relationship between the best-fit characteristic timescales, or periods, for the 34 events which preferred model S1 in both datasets. The events where model S1 was strongly preferred in both datasets are marked in red. These events are summarised in Table \ref{table2}. Here, the uncertainties associated with the data points are based on the best-fit parameter $\sigma$, which defines the breadth of the enhancement in frequency space. From Figure \ref{gbm_vs_goes}c, there is clearly a relationship between $P_{GBM}$ and $P_{GOES}$, and in most cases the detected timescales are consistent between the two instruments. Using a Spearman rank correlation test, we find that $\rho$ = 0.86 when only the strong S1 events are considered, and $\rho$ = 0.41 when all 34 events are considered. This reduction in correlation is primarily the result of three outlying events where $P_{GBM}$ is much larger than $P_{GOES}$. Figure \ref{gbm_vs_goes}d, further illustrates the relationship by plotting the ratio $P_{GBM} / P_{GOES}$, illustrating that most events are consistent with a ratio of 1. This consistency between the timescales detected by two instruments observing in different energy ranges give us confidence in the employed methodology, and also has implications as to the physical mechanisms occurring in these flares. Additionally, the persistence of this relationship in events that we have not identified as strongly favouring model S1 in Section \ref{goes_results} and Section \ref{gbm_results} suggests that these features may also be physical, and that the criteria adopted in this work are rather strict.

\begin{table*}[]
\centering
\caption{Summary of events where model $S_1$ is strongly favoured in both GOES and GBM datasets. Strongly favoured events in both instruments are marked with the $\dagger$ symbol.}
\label{table2}
\begin{tabular}{c|c|ccccc|ccccc}
                   &            &     \multicolumn{5}{c|}{GOES}                                               & \multicolumn{5}{c}{GBM}                  \\
                     \hline
GOES Start time (UT) & GOES Class & $\Delta$BIC$_{0-1}$ & $\Delta$BIC$_{2-1}$ & P (S) & $\sigma$  & $\alpha$ & $\Delta$BIC$_{0-1}$ & $\Delta$BIC$_{2-1}$  & P (s) & $\sigma$  & $\alpha$ \\
\hline
2011-03-07 07:59     &  M1.4      & 13.4          & 0.6           & 4.8          & 0.12        & 3.63            & 17.4         & 0.7          & 30.6        & 0.25       & 2.56           \\
2011-03-07 21:45     &  M1.5      & 3.2           & 7.9           & 15.5         & 0.07        & 4.21            & 3.9          & 9.9          & 13.7        & 0.18       & 2.52           \\
2011-03-08 02:24     &  M1.3      & 42.1          & 37.0          & 11.3         & 0.16        & 5.10            & 4.1          & 10.5         & 4.6         & 0.05       & 2.83           \\
2011-03-14 19:30$\dagger$ &  M4.2      & 179.9         & 79.6          & 8.2          & 0.12        & 4.22            & 56.9         & 43.5         & 8.1         & 0.05       & 3.32           \\
2011-09-23 23:48     &  M1.9      & 34.7          & 31.6          & 11.9         & 0.05        & 3.86            & 3.0          & 5.1          & 13.6        & 0.09       & 2.53           \\
2011-10-02 00:37     &  M3.9      & 14.9          & 10.0          & 36.7         & 0.25        & 4.10            & 33.4         & 9.1          & 45.2        & 0.22       & 3.52           \\
2011-11-06 06:14     &  M1.4      & 6.6           & 16.7          & 33.1         & 0.09        & 4.03            & 0.9          & 5.6          & 7.3         & 0.05       & 3.10           \\
2012-03-09 03:22$\dagger$     &  M6.3      & 74.9          & 57.6          & 65.8         & 0.14        & 3.50            & 27.9         & 27.6         & 69.3        & 0.15       & 2.82           \\
2012-03-13 17:12     &  M7.9      & 38.0          & 7.3           & 120.6        & 0.25        & 3.99            & 22.7         & 0.1          & 116.3       & 0.25       & 3.42           \\
2012-03-23 19:34     &  M1.0      & 43.1          & 17.4          & 10.8         & 0.25        & 4.49            & 8.2          & 2.2          & 11.8        & 0.25       & 4.42           \\
2012-07-04 09:47     &  M5.3      & 0.8           & 0.4           & 7.3          & 0.05        & 3.79            & 34.0         & 11.7         & 10          & 0.25       & 3.13           \\
2012-07-05 03:25     &  M4.7      & 27.3          & 8.9           & 8            & 0.05        & 4.14            & 5.9          & 2.1          & 5.2         & 0.11       & 3.68           \\
2012-10-22 18:38$\dagger$     &  M5.0      & 58.8          & 36.4          & 14           & 0.17        & 3.99            & 24.3         & 28.7         & 14.1        & 0.15       & 2.87           \\
2013-01-11 08:43     &  M1.2      & 8.7           & 14.5          & 44.1         & 0.07        & 3.88            & 10.2         & 11.0         & 45          & 0.08       & 2.92           \\
2013-04-11 06:55     &  M6.5      & 33.7          & 17.4          & 67.9         & 0.17        & 3.90            & 5.7          & 3.1          & 9.4         & 0.05       & 3.06           \\
2013-10-27 12:36     &  M3.5      & 34.7          & 27.5          & 16.8         & 0.05        & 3.59            & 0.3          & 5.8          & 16.7        & 0.05       & 3.23           \\
2013-11-05 08:12     &  M2.5      & 13.0          & 11.2          & 7.6          & 0.11        & 3.74            & 12.2         & 5.0          & 6.9         & 0.20       & 3.34           \\
2013-11-05 18:08     &  M1.0      & 47.7          & 17.0          & 14.3         & 0.25        & 5.06            & 22.4         & 7.2          & 9.3         & 0.12       & 2.58           \\
2013-11-07 03:34$\dagger$     &  M2.3      & 46.3          & 17.5          & 13.8         & 0.25        & 4.91            & 12.2         & 12.5         & 10.8        & 0.10       & 2.32           \\
2013-12-29 07:49     &  M3.1      & 14.5          & 13.8          & 13.9         & 0.25        & 4.16            & 7.9          & 8.6          & 15.4        & 0.17       & 3.04           \\
2014-01-28 11:34     &  M1.4      & 9.4           & 5.3           & 9.5          & 0.05        & 4.98            & 71.8         & 39.3         & 10          & 0.09       & 2.64           \\
2014-02-14 12:29     &  M1.6      & 15.2          & 14.1          & 18.1         & 0.25        & 3.65            & 6.2          & 0.1          & 16.4        & 0.17       & 2.15           \\
2014-03-12 22:28$\dagger$     &  M9.3      & 39.1          & 38.0          & 13.2         & 0.25        & 4.40            & 21.7         & 10.7         & 11.1        & 0.25       & 3.62           \\
2014-05-08 09:59     &  M5.2      & 51.9          & 40.6          & 27.3         & 0.25        & 4.98            & 27.3         & 5.1          & 26.6        & 0.25       & 2.65           \\
2014-06-06 19:26$\dagger$     &  M1.4      & 13.9          & 14.0          & 8.6          & 0.25        & 4.16            & 11.7         & 15.9         & 7.8         & 0.13       & 2.95           \\
2014-06-12 09:23     &  M1.8      & 18.5          & 15.1          & 7.4          & 0.24        & 3.67            & 2.4          & 3.9          & 25.9        & 0.17       & 2.32           \\
2014-10-09 01:30     &  M1.3      & 45.7          & 16.1          & 13.7         & 0.25        & 4.51            & 3.7          & 3.2          & 10.7        & 0.25       & 3.83           \\
2014-10-09 06:48     &  M1.2      & 13.1          & 8.2           & 30.2         & 0.16        & 3.83            & 0.3          & 1.9          & 37.7        & 0.25       & 2.67           \\
2014-10-29 21:18$\dagger$     &  M2.3      & 74.9          & 65.9          & 7.1          & 0.13        & 4.37            & 22.7         & 11.4         & 7.9         & 0.17       & 3.15           \\
2015-03-12 21:44     &  M2.7      & 12.9          & 14.9          & 15           & 0.07        & 3.88            & 4.0          & 10.5         & 28.3        & 0.23       & 2.59           \\
2015-05-05 13:45     &  M1.2      & 2.5           & 0.8           & 19           & 0.06        & 3.82            & 0.1          & 2.3          & 23.2        & 0.11       & 2.56           \\
2015-05-06 11:45     &  M1.9      & 11.6          & 0.9           & 237.7        & 0.25        & 3.01            & 29.1         & 21.4         & 4.9         & 0.05       & 3.12           \\
2015-06-11 08:49     &  M1.0      & 65.6          & 4.2           & 10.8         & 0.25        & 5.50            & 16.6         & 13.7         & 9.5         & 0.10       & 3.32           \\
2015-09-29 05:05     &  M2.9      & 43.1          & 27.6          & 27.2         & 0.25        & 4.73            & 12.8         & 4.2          & 21.4        & 0.25       & 2.74          
\end{tabular}
\end{table*}

\subsection{Further data exploration}
\label{data_exploration}

Using these datasets, a number of additional relationships can be tested. In particular, it is important to understand whether the characteristic timescale is dependent on any other measurable parameters of solar flares, such as the duration or the GOES-class of the event. We can also establish whether the measured power-law index of the Fourier power spectrum is related to these parameters; previous work has suggested that the power-law index may be a function of observation energy \citep{2007ApJ...662..691M, 2015ApJ...798..108I}.

\begin{figure*}[ht]
\begin{center}
\includegraphics[width=18cm]{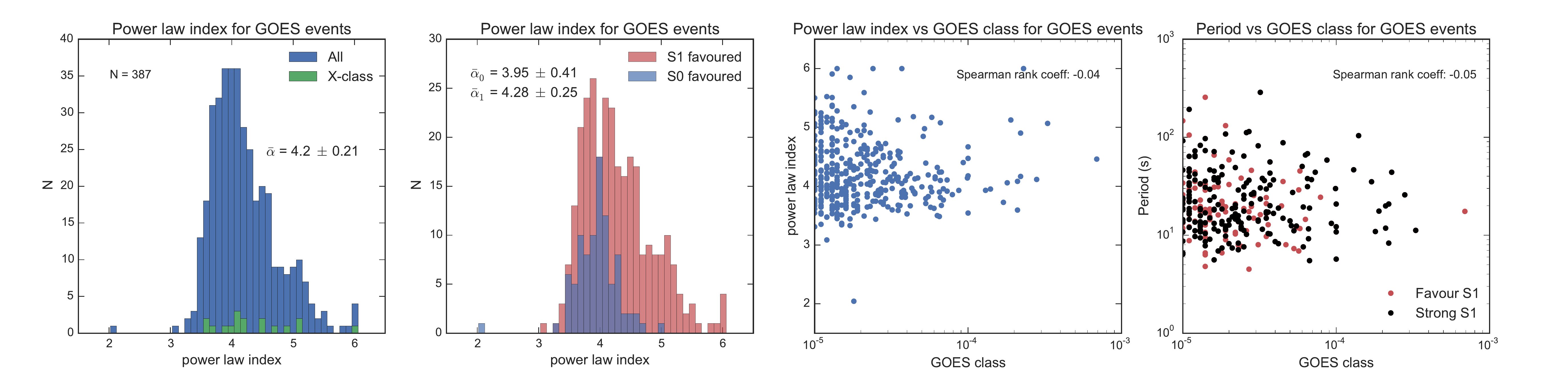}
\includegraphics[width=18cm]{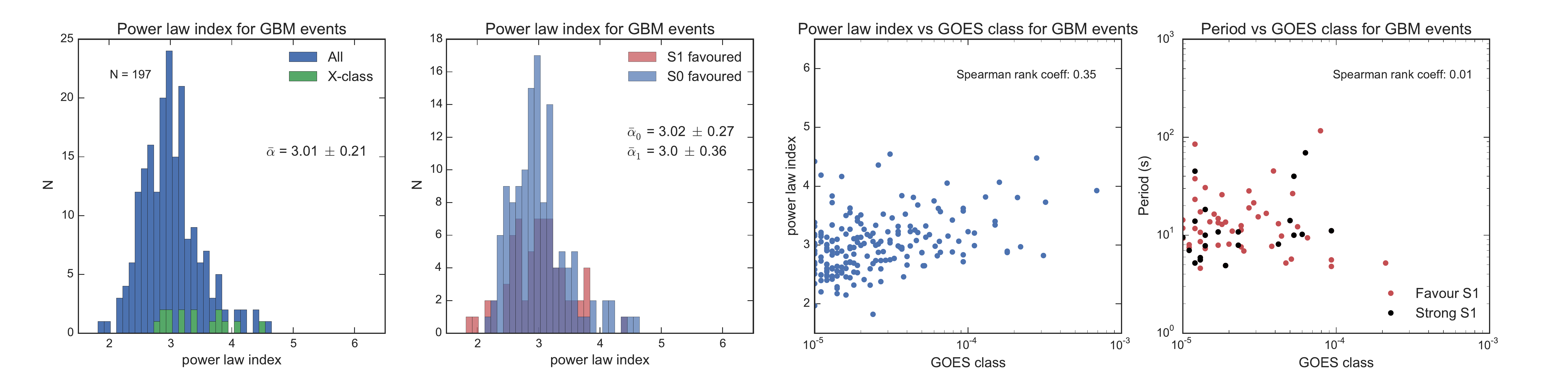}
\caption{Data exploration for GOES and GBM data sets. Left panels: histograms of best-fit power law index for all events where $S_0$ or $S_1$ was favoured and with an acceptable goodness-of-fit, observed by GOES (top) and GBM (bottom). M-class events are shown in blue while X-class flares are highlighted in green. Center-left panels: Histograms of power-law index for GOES (top) and GBM (bottom) as before, separated into categories according to whether $S_0$ or $S_1$ was favoured. Center-right panels: Best-fit power law index vs GOES-class for GOES (top) and GBM (bottom) events. We find Spearman rank correlation coefficients of $C_{GOES}$ = -0.04 and $C_{GBM}$ = 0.35, indicating little correlation in GOES and weak correlation in GBM. Right panels: Best-fit period vs GOES-class for all events strongly favouring model $S_1$ (black data) and favouring $S_1$ by any margin (red data), for GOES (top) and GBM (bottom). The Spearman rank correlation coefficients are $C_{GOES}$ = -0.05 and $C_{GBM}$ = 0.01, indicating no correlation.}
\label{data_exploration_fig}
\end{center}
\end{figure*}

\begin{figure*}
\begin{center}
\includegraphics[width=18cm]{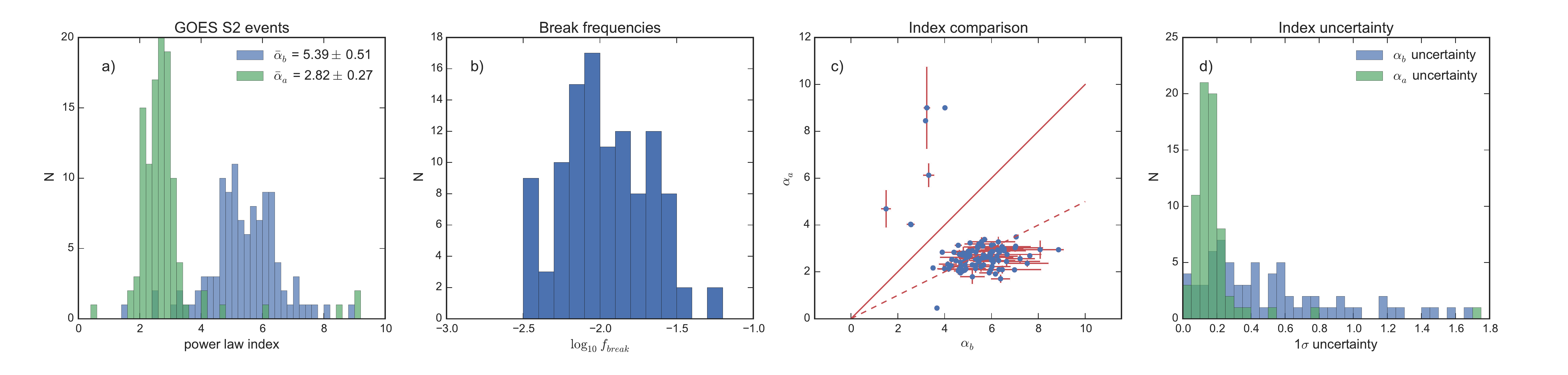}
\caption{Properties of the best-fit model for flares classified as S2 in GOES data. a) Histogram of the power law indices $\alpha_b$ (below the break frequency $f_{break}$) and $\alpha_b$ (above $f_{break}$). b) Distribution of break frequencies $f_{break}$. c) $\alpha_a$ vs $\alpha_b$ for all S2 events. The solid red line denotes $y$ = $x$. The dashed line denotes $y$ = $x/2$. In almost all cases, the spectral index flattens above the break frequency. d) Distribution of the uncertainties in $\alpha_a$ and $\alpha_b$.  }
\label{s2_events_goes}
\end{center}
\end{figure*}

In Figure \ref{data_exploration_fig} we show the distributions of the best-fit power-law index for both GOES and GBM datasets, when either model $S_0$ or $S_1$ is favoured. Events where $S_2$ is preferred are considered separately in Figure \ref{s2_events_goes}. For these distributions we also remove from consideration any events where the preferred model fails the goodness-of-fit test described in Section \ref{method}. The left panels show the distribution of power-law indices for all these events, separated into M- and X-class. We can see that the mean of the distribution is rather different for the two instruments; the mean power-law index and standard error for GOES events is $\bar{\alpha}_{GOES} $ = 4.20 $\pm$ 0.21 whereas for GBM the mean is $\bar{\alpha}_{GBM}$ = 3.01 $\pm$ 0.21. Some GOES events show power law indices as high as 6, while no GBM events show an index steeper than 4.5. In both cases, there is no obvious difference between M- and X-class events. The second column of Figure \ref{data_exploration_fig} shows these distributions separated into two categories based on the favoured model. For GOES we find that $\bar{\alpha}_0 $ = 3.95 $\pm$ 0.41, and $\bar{\alpha}_1$ = 4.28 $\pm$ 0.25, indicating that although the measured mean is slightly steeper for events preferring $S_1$, the means are consistent within the uncertainties. This suggests that there is no strong selection effect occurring. For example, steeper overall Fourier power spectra may allow a relatively modest signal at higher frequencies to be discovered, whereas a flatter spectrum may mask such a component. The consistency of $\bar{\alpha}_0$ and $\bar{\alpha}_1$ suggests that this is not the case for our sample. Similarly, in the GBM data there we find $\bar{\alpha}_0$ = 3.02 $\pm$ 0.27 and $\bar{\alpha}_1$ = 3.00 $\pm$ 0.36, consistent within the error bar. The panels in the third column show the relationship between the measured power-law indices and the GOES-class of each event. For GOES, we see that there is little obvious relationship between the two. Using the Spearman rank correlation test, we find coefficients of $C_{GOES}$ = -0.04. For GBM there appears to be a slight relationship with $C_{GBM}$ = 0.35, and steeper indices preferentially associated with larger GOES class. Finally, the fourth column shows the relationship between the detected period and the GOES class for GOES (top) and GBM (bottom) data, for all events where model $S_1$ is strongly favoured (black), and where it is favoured by any margin (red). Here in both cases we see Spearman correlation coefficients close to zero, indicating that there is no relationship between the flare size and the detected period for either instrument.

We performed several other correlation tests which for brevity are not shown in Figure \ref{data_exploration_fig}. In particular, we investigated the relationship between $P$ and the event duration, $P$ and $\Delta$BIC, and GOES-class and $\Delta$BIC. In each case we found no evidence of a correlation in either dataset. From this we can conclude that the flare size and duration are not major factors that determine QPP properties or their occurrence. 

\citet{2007ApJ...662..691M, 2015ApJ...798..108I} have previously suggested a relationship between the measured power-law index in Fourier power spectra and energy of the emission, in the context of self-organised criticality and fractal behaviour. In particular, \citet{2007ApJ...662..691M} calculated the Holder exponent for emission at various energies for the 2002 July 3 solar flare, a property which is related to the Fourier power-law index. They relate the flattening of this index with energy to the persistence - i.e., how later observed values in a signal depend on earlier values of the physical system. In a `persistent' regime, if an increase in the signal is observed then the next observed value in time is also likely to show an increase (smooth behavior), whereas in an anti-persistent regime the opposite occurs, and a decrease is equally likely (bursty behavior). The results presented in Figure \ref{data_exploration_fig} are consistent with this picture, with the 15-25 keV emission showing a systematically flatter mean power-law index of $\sim$ 3, compared to the lower energy 1-8 \AA ($\sim$ 1.5 - 12 keV) GOES emission with a mean index of $\sim$ 4.2. This is also consistent with our understanding of flare properties, where the lower energy emission is related to thermal heating, where the cooling time is long relative to the electron beam decay time, whereas higher energy emission is increasingly associated with burstier impulsive particle acceleration and energy deposition.

Also of interest are the properties of events where the broken-power law model $S_2$ is preferred, for example the typical values of the power law indices $\alpha_a$, $\alpha_b$ and the break frequency $f_{break}$. Figure \ref{s2_events_goes} examines these properties for the 110 events classed as S2 in GOES data. We can see that the distributions of $\alpha_b$ and $\alpha_a$, the power-law indices below and above the break frequency $f_{break}$, are very different, with a mean value of $\bar{\alpha}_b$ = 5.39 $\pm$ 0.51 for below the break and $\bar{\alpha}_a$ = 2.82 $\pm$ 0.27 above. Hence, the indices below the break frequency are generally much steeper than those above the break frequency. The value of $f_{break}$ itself is distributed around a peak value of $\sim$ 0.01 Hz, corresponding to $P$ $\sim$ 100 s. Figure \ref{s2_events_goes}c shows $\alpha_a$ plotted directly against $\alpha_b$ for all GOES S2 events. From this it can be seen that in all but a few occasions, the spectral index flattens above the break frequency, in general by a factor of $\sim$ 2. Figure \ref{s2_events_goes}d shows the distribution of the 1$\sigma$ uncertainties for $\alpha_a$ and $\alpha_b$, which for figure clarity we omit from the previous panel. We can see that the value of $\alpha_a$ is typically well-constrained to within $\sim$ 0.2, whereas a greater spread of uncertainty is associated with the $\alpha_b$ values. This is due to the relative scarcity of points below $f_{break}$ in these cases. For the outlying points above the $y$=$x$ line in Figure \ref{s2_events_goes}d, the uncertainties are small enough that these points are clearly separated from the rest of the population.

\begin{figure}
\begin{center}
\includegraphics[width=7cm]{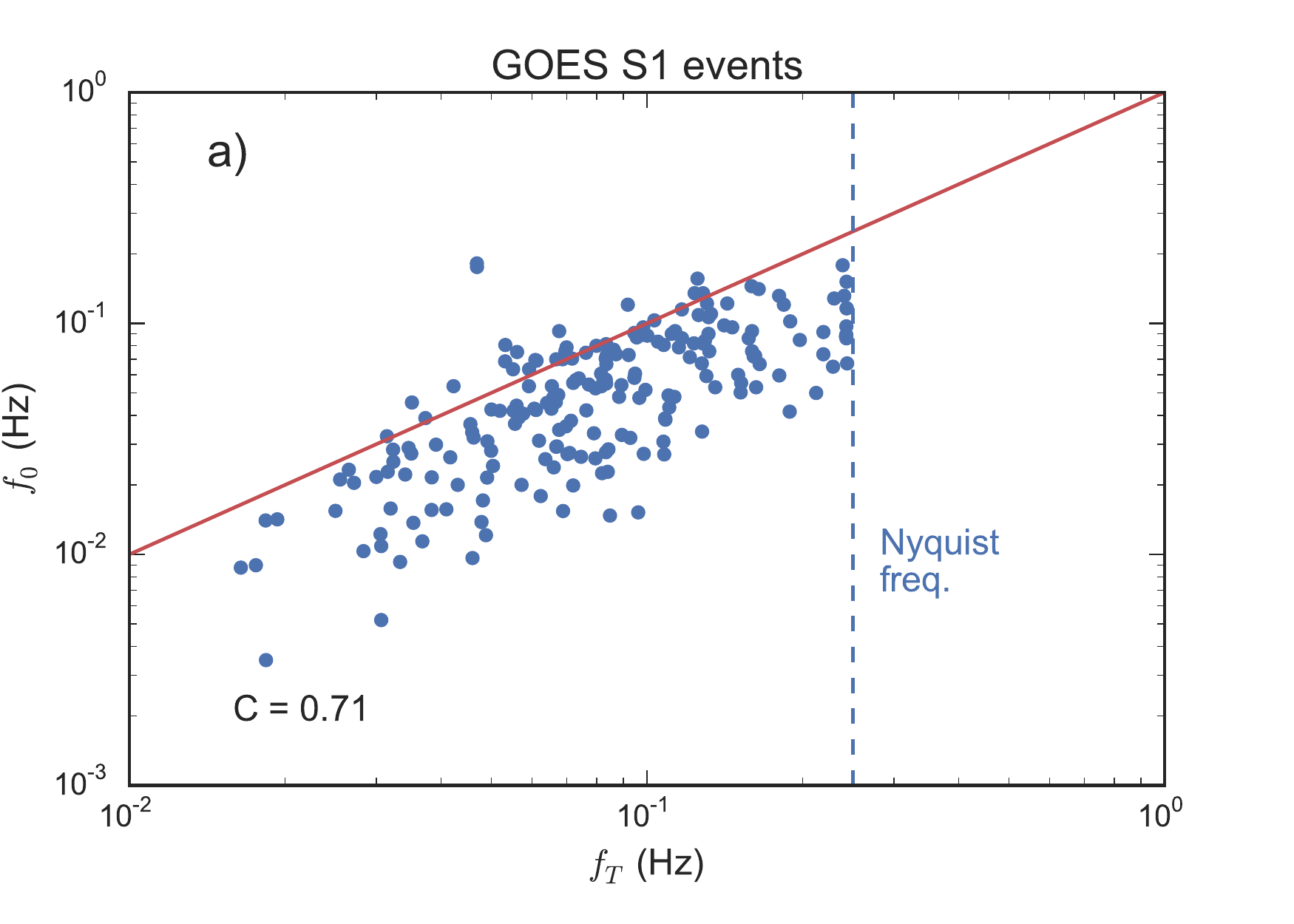}
\includegraphics[width=7cm]{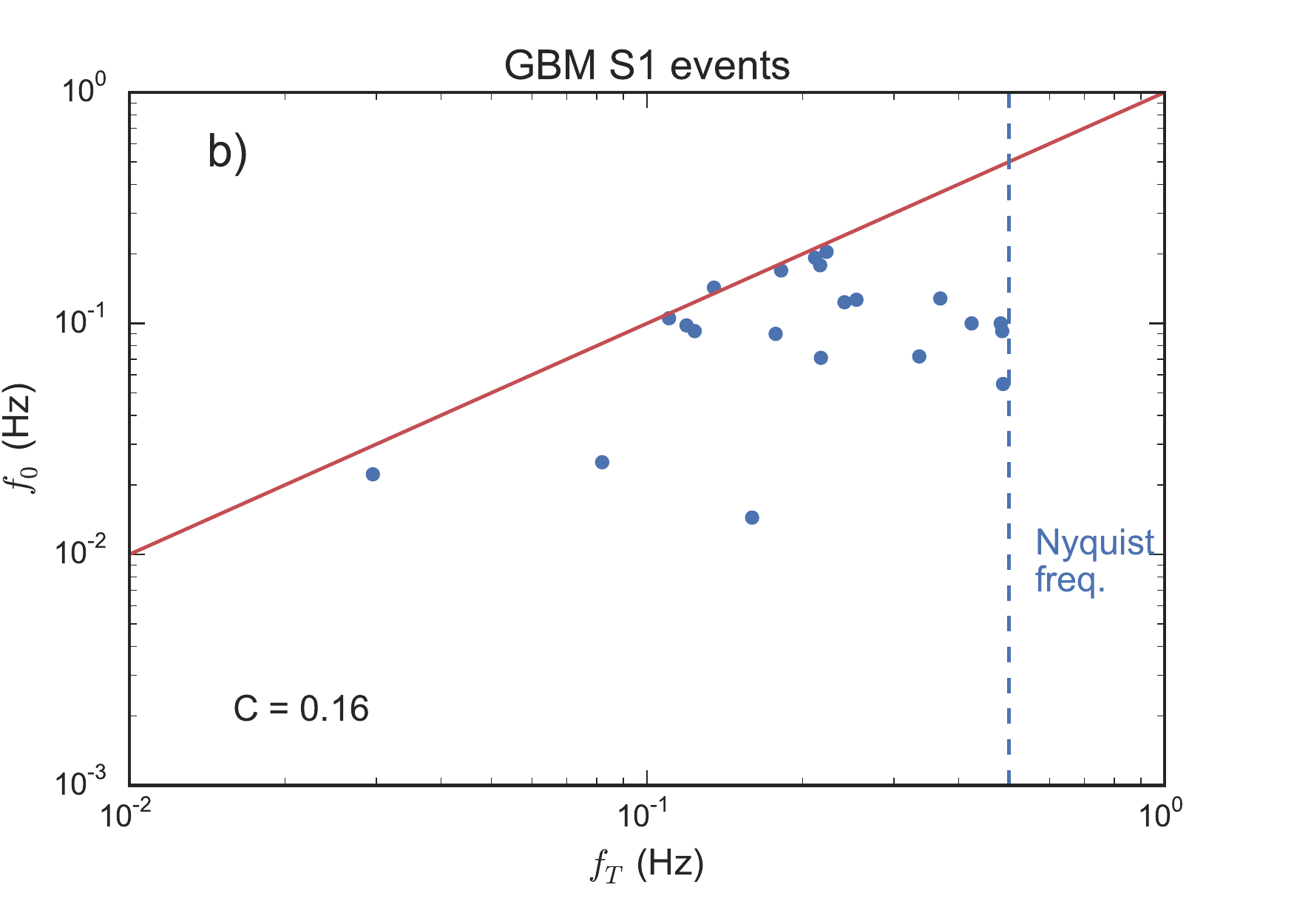}
\caption{Correlation between the best-fit frequency $f_0$ and the turnover frequency $f_T$ for S1 events in a) GOES data and b) GBM data. The Spearman rank correlation in the GOES dataset is $C_{GOES}$ = 0.71, while for GBM we find $C_{GBM}$ = 0.16. The red line in each panel represents $f_0 = f_T$.}
\label{period_vs_ft}
\end{center}
\end{figure}

In Figure \ref{period_vs_ft} we examine the relationship between \textbf{best-fit frequency $f_0 = 1/P$ and} the turnover frequency $f_T$, which denotes the transition between a power-law and a white-noise regime in the Fourier power spectrum. We can see that, at least in the GOES dataset, there is a correlation between the turnover and the best-fit period. We find Spearman rank correlation coefficients of $C_{GOES}$ = 0.71 and $C_{GBM}$ = 0.16. Additionally, it can be seen that in both datasets it is rare for $f_0$ to be found at a higher frequency than $f_T$. This is illustrated in Figure \ref{period_vs_ft} by the solid red line, which represents $f_0 = f_T$ - clearly, the majority of the best-fit $f_0$ values lie below this line. Hence, in general, characteristic periods tend to be found in the power-law dominated part of the Fourier spectrum. This can be explained by the fact that the portion of the Fourier power spectrum above $f_T$ is dominated by background noise, rather than flare signal (see examples in Figures \ref{goes_examples}, \ref{gbm_examples}). Therefore, any weak oscillatory signals in this regime are likely to remain undetected by the model comparison method. The weaker correlation in the GBM dataset can be explained both by the small sample size and by the fact that the distribution of $f_T$ values is peaked at a higher frequency than in GOES data (see Figure \ref{afino_distributions_gbm}e). 

\section{Discussion and Interpretation}
\label{discussion}

We have analysed 675 GOES events and 261 GBM solar flares using a model comparison approach similar to that introduced by \citet{2015ApJ...798..108I}. The purpose of this analysis is to identify events with signatures consistent with QPP. We considered three models; a single power-law-plus-constant model $S_0$ (see Equation \ref{m0_eqn}), a broken power-law model $S_2$ (see Equation \ref{m2_eqn}) and a model $S_1$ including a power-law but also an additional peak in the Fourier power spectrum (see Equation \ref{m1_eqn}). All three of these models account for the known power-law shape of flare-like signals \citep{2010AJ....140..224C, 2011A&A...533A..61G, 2013ApJ...768...87H, 2015ApJ...798..108I, 2015ApJ...798....1I}. Choosing the Bayesian Information Criterion as our model selection statistic, we search for events with a strong preference for model $S_1$, defined as $BIC_1$ < $BIC_{j}$ - 10 for $j$=0,2. We find that $S_1$ is a more appropriate fit in 202 of 675 GOES events (30\%), and in 21 out of 261 GBM events (8\%, see Figures \ref{afino_distributions}, \ref{afino_distributions_gbm}). It is these signatures that we consider most consistent with classical ideas of QPP, namely that they are a signature of MHD wave modes, or quasi-periodic magnetic reconnection. This suggests that QPP signals are a relatively uncommon occurrence in solar flares, or at least of weak amplitude relative to the global Fourier power spectrum.

There are a number of caveats to consider when interpreting this estimate of the prevalence of such signals. Firstly, the method employed here is a global technique; we analyse the entire Fourier power spectrum for the duration of the event, rather than employing a time-dependent method such as a wavelet analysis. Hence, if a signal is only present during a relatively small fraction of the flare, its overall strength in the Fourier power spectrum may be small enough that it goes undetected by our model comparison approach. Additionally, phase changes of an oscillation during an event may also cause a signal to remain undetected.

A second drawback is the scenario where the characteristic timescale is rapidly evolving during an event. For example, recent work \citep[see][]{2016Dennis} has shown that in one flare the characteristic timescale evolved from $\sim$ 20 s to $\sim$ 80 s during one hour of observations following the GOES event peak. Using our global technique, the result of this would be a very broad peak in Fourier space. Depending on the location of this peak in $f$, such a feature may not be fully captured by model $S_1$ due to the width limitations imposed in order to ensure that a localized feature is detected (see Equation \ref{limits}). Time-dependent methods, such as wavelet analysis \citep[see e.g.][]{2015SoPh..290.3625S, 2016Dennis} may be better suited for exploring this category of events. If such events are common it would substantially change the estimated prevalence of QPP signatures, although regardless the result presented here is much lower than that suggested recently by \citet{2015SoPh..290.3625S}, who found an $\sim$ 80\% occurrence rate of QPP in X-class flares.

In order to obtain statistically robust, reproducible results, we have attempted to be as consistent as possible in the analysis of both the GOES and Fermi/GBM data. To achieve this, it was inevitably necessary to impose some constraints on the analysis method and sacrifice some ability to customize the data selection. For example, in order to have a consistent methodology for the selection of the analysis time interval, we used the start and end times recorded in the GOES catalogue. In reality, when analysing a single event, it may be desirable to fine-tune this interval in order to best capture the perceived time interval of interest. Similarly, for analysing Fermi/GBM data we prescribe the end time to be midway between the GOES peak time and GOES end time, regardless of the true shape of the Fermi/GBM lightcurve. Again, for analysing a small sample in detail, such parameters could be adjusted as needed to suit the individual circumstances. For this study however, it was critical to maintain a consistent approach in order for the results to be reproducible. 

An additional finding of this work is that, for both instruments, the detected timescales tended to be clustered in a distinct range. In both cases, the period distributions are centered at $\sim$ 10-15s (although events with $P < 10 s$ may have been missed, particularly in GOES data), with most events falling in the $P < 50$ s range (see Figures \ref{afino_distributions}, \ref{afino_distributions_gbm}). This is meaningful since this is consistent with the expected characteristic periods of various MHD modes in the solar corona \citep[e.g.][]{2008A&A...481..819M, 2009A&A...494.1119P, 2009A&A...503..569I, 2011A&A...526A..75M, 2014ApJ...784..101P}, see also \citet{2009SSRv..149..119N, 2012RSPTA.370.3193D, 2016SSRv..200...75N} for recent reviews. This preferred timescale also suggests that there is a physical feature, e.g. a length scale or time scale, that is common to flares exhibiting QPP. In Section \ref{data_exploration} no relationship between flare magnitude and $P$ was found (see Figure \ref{data_exploration_fig}), meaning that the amount of energy released is not a major factor that determines QPP properties or occurrence probability. A similar result was found for stellar flare QPP observed by Kepler \citep{2016MNRAS.tmp..660P}. This suggests in turn that the physical size of the flare arcade may not be critical, as more energetic events tend to be associated with larger arcades. 

It also remains a challenge to explain QPP in flares purely in terms of MHD wave modes given the complex motion and evolution of flare plasma, in particular the movement of X-ray emission footpoints \citep[e.g.][]{2005ApJ...625L.143G, 2005AdSpR..35.1707K, 2009ApJ...693..132Y, 2013ApJ...777...30I}. Another possible explanation for QPP signatures is the generation and interaction of magnetic islands in flare current sheets \citep[e.g.][]{2006Natur.443..553D, 2013ApJ...763L...5D, 2016ApJ...820...60G}, which may explain the bursty generation of high-energy electrons and hard X-ray emission. In this scenario, the magnetic islands can trap and accelerate electrons, increasing their energy by two orders of magnitude \citep{2006Natur.443..553D}. As the islands descend sunward and interact with the flare arcade, these electrons escape, streaming to the flare footpoints and generating X-ray emission \citep{2016ApJ...820...60G}. Hence, the preferred timescale of $\sim$ 5-30 s found in this work provides a constraint on the rate of island formation in the current sheet. An advantage of this model is that it self-consistently explains the observed motions of hard X-ray footpoints in the chromosphere as well as the steady increase in height of hot plasma in flare arcades over time \citep[e.g.][]{2013ApJ...767..168L}.

\section{Summary}

We have performed a large-sample search for quasi-periodic pulsations in solar flares over the time interval 2011 February 1 - 2015 December 31, analysing 675 GOES events and 261 Fermi/GBM events. To the best of our knowledge, this is the largest statistical study focused on QPP in solar flares. Our main findings are as follows:

\begin{itemize}
\item Approximately 30\% of GOES events (202), and 8\% of GBM events (21) showed strong evidence of an enhancement in the Fourier power spectrum consistent with classical ideas of QPP.

\item When the model selection criterion (see Equation \ref{strong_test}) was relaxed to $\Delta$BIC > 0, a total of 295 GOES events (44\%) and 69 GBM events (26\%) preferred the QPP-like model $S_1$.

\item For both instruments, characteristic timescales in the 5-30s range were most commonly detected, as shown in Figures \ref{afino_distributions}f, \ref{afino_distributions_gbm}f).

\item For the 7 events where model $S_1$ was strongly preferred in both GBM and GOES, we find a correlation between the best-fit periods for each instrument (see Figure \ref{gbm_vs_goes}). If all 34 events are included where $S_1$ is favoured by any margin in both instruments, this correlation is replicated.  This indicates that the same features and timescales are being seen in two energy bands, 1-8\AA\ and 15-25 keV.

\item Higher energy 15-25 keV GBM data shows on average shallower Fourier power-law indices than the GOES 1-8 \AA\ data, with a mean value of $\bar{\alpha}_{GBM}$ = 3.02 $\pm$ 0.24 in GBM, compared to $\bar{\alpha}_{GOES}$ = 4.20 $\pm$ 0.21 in GOES.

\item No significant correlations were found in the GOES dataset between the detected period and the flare size (GOES class), or the best-fit power-law index and flare size. A weak correlation between $P$ and flare size was found in the GBM data. This suggesting that the flare magnitude does not have a major impact on QPP properties.
\end{itemize}

Hence, we can draw preliminary conclusions about the occurrence rate of QPP-like signatures in solar flares, which we conservatively estimate to be $\geq$ 30\% in Soft X-rays and $\geq$ 8\% in hard X-rays. These results indicate that QPP signatures are a minority occurrence in solar flare data. However, as discussed in Section \ref{discussion}, the true rate may be higher, as we have adopted conservative criteria in this study. Although this work does not attempt to determine the mechanism responsible for producing the observed signatures, the finding of a characteristic timescale of 5-30 s is informative, as such timescales are consistent with the expected signature of MHD wave modes in flare arcades. Much scope for future work remains - in particular statistical surveys of QPP in flares should be performed for other instruments at different wavelengths in order to establish a complete picture of QPP occurrence in solar flares. Similar studies can also be carried out on stellar flare data \citep[e.g.][]{2015MNRAS.450..956B}, while further study of the nature of events where a broad feature in the Fourier power spectrum was observed is also needed. Finally, future work is needed to explore the differences in occurrence and evolution of QPP between the impulsive and gradual phases of solar flares.

\appendix
The large set of results collected by this study are useful as a guide for future work on QPP. In Tables \ref{goes_list} and \ref{gbm_list} we present excerpts of the list of obtained parameters for the 675 GOES flares and 261 GBM flares we have analysed. These tables are published in full in electronic format.

\begin{table*}[]
\centering
\caption{Abbreviated list of studied GOES events and analysis result parameters. The full table is available electronically.}
\label{goes_list}
\begin{tabular}{c|cccccccccccccccc}
 & GOES & Start & End & & & & & & & & & & &   \\
Date     & class & time & time & $\Delta$BIC$_{0-1}$ & $\Delta$BIC$_{0-2}$ & $\Delta$BIC$_{2-1}$ & $\chi^2_{0}$ & $p_{0}$ & $\chi^2_{1}$ & $p_1$ & $\chi^2_2$ & $p_2$ & Period (s) & $\sigma$             \\
\hline
20110209 & M1.9       & 012301      & 013458      & 15.4                & 8.6                 & 6.8                 & 1.13         & 0.26    & 0.92         & 0.62  & 0.95       & 0.56  & 11.1       & 0.25        \\
20110213 & M6.6       & 172801     & 174659       & 18.7                & 8.1                 & 10.6                & 1.21         & 0.11    & 1.12         & 0.24  & 1.09       & 0.28  & 11.6       & 0.25          \\
20110214 & M2.2       & 172001     & 173158       & 36.2                & 27.2                & 9.0                 & 1.15         & 0.23    & 1.02         & 0.44  & 0.92       & 0.62  & 10.7       & 0.25           \\
20110215 & X2.2       & 014401      & 020558      & 104.6               & 45.5                & 59.1                & 1.68         & 0.00    & 1.28         & 0.05  & 1.48       & 0.00  & 20.8       & 0.25         \\
20110216 & M1.0       & 013200      & 014558      & 10.4                & 12.0                & -1.6                & 0.93         & 0.61    & 0.80         & 0.84  & 0.89       & 0.69  &       &                 \\
20110216 & M1.1       & 073500      & 075458      & -1.5                & 12.8                & -14.3               & 1.01         & 0.46    & 0.96         & 0.58  & 0.87       & 0.77  &            &               \\
20110216 & M1.6       & 141900     & 142858       & -9.9                & -8.1                & -1.8                & 0.59         & 0.98    & 0.54         & 0.99  & 0.55       & 0.99  &            &               \\
20110218 & M6.6       & 095501      & 101459      & 116.8               & 79.5                & 37.3                & 1.03         & 0.40    & 0.71         & 0.97  & 0.95       & 0.60  & 9.2        & 0.25           \\
20110218 & M1.0       & 102300     & 103657       & -10.6               & -9.2                & -1.4                & 0.80         & 0.84    & 0.76         & 0.90  & 0.79       & 0.86  &            &               \\
20110218 & M1.4       & 125901     & 130559       & -4.5                & -2.6                & -1.9                & 0.88         & 0.63    & 0.74         & 0.82  & 0.75       & 0.81  &            &              \\
20110218 & M1.0       & 140001     & 141458       & 44.7                & 50.5                & -5.8                & 1.47         & 0.01    & 1.24         & 0.11  & 0.99       & 0.49  &        &                \\
20110218 & M1.3       & 205602     & 211359       & 35.0                & 14.4                & 20.7                & 1.37         & 0.03    & 1.03         & 0.41  & 1.27       & 0.07  & 18.2       & 0.25            \\
20110224 & M3.5       & 072301      & 074158      & 0.3                 & 0.0                 & 0.3                 & 1.06         & 0.35    & 0.94         & 0.62  & 0.97       & 0.55  & 16.6       & 0.05             \\
20110228 & M1.1       & 123801     & 130258       & 8.1                 & 27.4                & -19.3               & 1.38         & 0.01    & 1.20         & 0.09  & 1.15       & 0.16  &          &                  \\
20110307 & M1.2       & 050001      & 051859      & 2.4                 & 2.5                 & -0.1                & 1.02         & 0.43    & 0.88         & 0.75  & 0.94       & 0.63  &      &                   \\
20110307 & M1.5       & 074900      & 075558      & 9.1                 & 23.3                & -14.2               & 1.12         & 0.30    & 1.09         & 0.34  & 0.93       & 0.56  &     &                 \\
20110307 & M1.4       & 075900      & 081459      & 13.4                & 12.8                & 0.6                 & 0.92         & 0.65    & 0.69         & 0.97  & 0.75       & 0.92  & 4.8        & 0.12          \\
20110307 & M1.8       & 091400      & 092759      & 4.0                 & 1.8                 & 2.2                 & 1.08         & 0.32    & 1.00         & 0.47  & 0.98       & 0.51  & 22.2       & 0.25          \\
20110307 & M1.9       & 134501     & 145558       & -5.5                & -7.2                & 1.7                 & 0.97         & 0.62    & 0.95         & 0.69  & 0.95       & 0.69  &            &           \\
20110307 & M3.7       & 194300     & 205759       & -19.8               & -11.2               & -8.6                & 3.30         & 0.00    & 3.31         & 0.00  & 3.34       & 0.00  &            &           \\
20110307 & M1.5       & 214501     & 215459       & 3.2                 & -4.6                & 7.9                 & 1.05         & 0.39    & 0.96         & 0.54  & 1.02       & 0.43  & 15.5       & 0.07             \\
20110308 & M1.3       & 022401      & 023158      & 42.1                & 5.2                 & 37.0                & 1.13         & 0.29    & 0.70         & 0.87  & 1.11       & 0.32  & 11.3       & 0.16             \\
20110308 & M1.5       & 033700      & 041958      & -5.9                & -1.8                & -4.2                & 1.40         & 0.00    & 1.40         & 0.00  & 1.40       & 0.00  &            &           \\
20110308 & M5.3       & 103501     & 105459       & 20.3                & 6.3                 & 14.0                & 1.48         & 0.00    & 1.19         & 0.13  & 1.25       & 0.07  & 11.1       & 0.10       \\
20110308 & M4.4       & 180800     & 184058       & 9.1                 & 11.1                & -2.1                & 1.69         & 0.00    & 1.65         & 0.00  & 1.57       & 0.00  &        &      	         \\
20110308 & M1.4       & 194602     & 211858       & -16.5               & -12.0               & -4.5                & 1.06         & 0.23    & 1.06         & 0.23  & 1.06       & 0.23  &            &                 \\
20110309 & M1.7       & 103501     & 112059       & 30.8                & 54.1                & -23.3               & 1.02         & 0.41    & 0.94         & 0.71  & 0.89       & 0.84  &        &                  \\
20110309 & M1.7       & 131700     & 140628       & 74.8                & 10.4                & 64.5                & 0.92         & 0.77    & 0.79         & 0.98  & 0.84       & 0.94  & 38.4       & 0.13           \\
20110309 & X1.5       & 231300     & 232859       & 6.4                 & 24.3                & -17.9               & 1.56         & 0.00    & 1.57         & 0.00  & 1.23       & 0.12  &        &           \\
20110310 & M1.1       & 223401     & 224858       & 83.4                & 68.1                & 15.3                & 1.41         & 0.02    & 0.97         & 0.54  & 1.09       & 0.31  & 14.5       & 0.25       
\end{tabular}
\begin{tablenotes}
 \item Note: Table 3 is published in its entirety in a machine readable format. A portion is shown here for guidance regarding its form and content.
\end{tablenotes}
\end{table*}

\begin{table*}[]
\centering
\caption{Abbreviated list of studied Fermi/GBM events and analysis result parameters. The full table is available electronically.}
\label{gbm_list}
\begin{tabular}{c|cccccccccccccc}
         & GOES  & Start & End &           &           &           &           &                 &           &                 &           &                 &        &  \\
Date     & class & time & time & $\Delta$BIC$_{0-1}$ & $\Delta$BIC$_{0-2}$ & $\Delta$BIC$_{2-1}$ & $\chi^2_{0}$ & $p_{0}$ & $\chi^2_{1}$ & $p_1$ & $\chi^2_2$ & $p_2$ & Period (s) & $\sigma$             \\
\hline
20110213 & M6.6       & 173009     & 174228   & -6.1      & -6.3      & 0.2       & 0.83      & 0.87            & 0.83      & 0.88            & 0.81      & 0.90            &        &       \\
20110214 & M2.2       & 172223     & 172859   & -6.7      & -8.6      & 1.9       & 0.79      & 0.85            & 0.76      & 0.89            & 0.80      & 0.84            &        &       \\
20110215 & X2.2       & 14603      & 20058    & -2.8      & -4.6      & 1.8       & 1.23      & 0.06            & 1.20      & 0.08            & 1.17      & 0.11            &        &       \\
20110216 & M1.0       & 13421      & 14229    & -13.0     & -8.2      & -4.8      & 0.80      & 0.87            & 0.80      & 0.86            & 0.80      & 0.86            &        &       \\
20110216 & M1.1       & 73715      & 74929    & -3.1      & 3.0       & -6.1      & 0.99      & 0.51            & 0.98      & 0.53            & 0.97      & 0.56            &        &       \\
20110216 & M1.6       & 142120     & 142659   & -11.6     & -7.3      & -4.3      & 0.88      & 0.69            & 0.88      & 0.68            & 0.87      & 0.71            &        &       \\
20110224 & M3.5       & 72721      & 73829    & -7.0      & -9.9      & 2.9       & 0.91      & 0.71            & 0.90      & 0.74            & 0.90      & 0.72            &        &       \\
20110228 & M1.1       & 124516     & 125729   & 6.1       & -4.8      & 10.9      & 1.20      & 0.10            & 1.12      & 0.21            & 1.15      & 0.16            & 7.5    & 0.10  \\
20110307 & M1.5       & 74956      & 75459    & -2.6      & -5.0      & 2.3       & 0.85      & 0.72            & 0.82      & 0.77            & 0.84      & 0.74            &        &       \\
20110307 & M1.4       & 80439      & 81059    & 17.4      & 16.7      & 0.7       & 1.49      & 0.02            & 1.08      & 0.33            & 1.13      & 0.26            & 30.6   & 0.25  \\
20110307 & M1.8       & 91753      & 92359    & -12.0     & -8.3      & -3.7      & 0.69      & 0.94            & 0.65      & 0.96            & 0.71      & 0.93            &        &       \\
20110307 & M1.9       & 141928     & 144258   & -15.3     & -10.5     & -4.8      & 0.94      & 0.69            & 0.94      & 0.69            & 0.94      & 0.69            &        &       \\
20110307 & M3.7       & 200007     & 203458   & -10.7     & -9.0      & -1.7      & 0.99      & 0.54            & 0.96      & 0.66            & 0.98      & 0.58            &        &       \\
20110307 & M1.5       & 214517     & 215229   & 3.9       & -6.0      & 9.9       & 0.97      & 0.53            & 1.01      & 0.44            & 1.02      & 0.43            & 13.7   & 0.18  \\
20110308 & M1.3       & 22601      & 23028    & 4.1       & -6.4      & 10.5      & 0.83      & 0.74            & 0.52      & 0.99            & 0.75      & 0.84            & 4.6    & 0.05  \\
20110308 & M4.4       & 181843     & 183428   & 8.8       & 27.5      & -18.8     & 1.03      & 0.41            & 0.97      & 0.59            & 0.92      & 0.71            & 43.9   & 0.25  \\
20110309 & M1.7       & 104815     & 111359   & 12.2      & 1.3       & 11.0      & 0.96      & 0.64            & 0.96      & 0.66            & 0.95      & 0.69            & 10.8   & 0.19  \\
20110309 & X1.5       & 231652     & 232558   & -0.3      & -6.4      & 6.1       & 0.60      & 1.00            & 0.57      & 1.00            & 0.58      & 1.00            &        &       \\
20110312 & M1.3       & 43338      & 44529    & -9.3      & -3.4      & -5.9      & 0.83      & 0.88            & 0.82      & 0.89            & 0.82      & 0.89            &        &       \\
20110314 & M4.2       & 194817     & 195259   & 56.9      & 13.5      & 43.5      & 2.14      & 0.00            & 0.79      & 0.80            & 1.22      & 0.17            & 8.1    & 0.05  \\
20110315 & M1.0       & 1915       & 2259     & 0.4       & 1.8       & -1.3      & 0.82      & 0.73            & 0.70      & 0.86            & 0.75      & 0.81            & 4.5    & 0.11  \\
20110325 & M1.0       & 231149     & 232559   & -11.3     & -5.2      & -6.1      & 1.07      & 0.31            & 1.06      & 0.33            & 1.04      & 0.36            &        &       \\
20110528 & M1.1       & 214723     & 215529   & -3.6      & -3.3      & -0.2      & 0.87      & 0.74            & 0.82      & 0.83            & 0.84      & 0.80            &        &       \\
20110607 & M2.5       & 62107      & 64959    & -10.0     & 13.0      & -23.0     & 0.95      & 0.68            & 0.96      & 0.66            & 0.93      & 0.77            &        &       \\
20110614 & M1.3       & 213814     & 215828   & -9.5      & -10.2     & 0.7       & 0.80      & 0.97            & 0.80      & 0.97            & 0.79      & 0.97            &        &       \\
20110730 & M9.3       & 20520      & 21029    & 0.3       & -2.4      & 2.7       & 0.85      & 0.72            & 0.77      & 0.83            & 0.73      & 0.88            & 4.8    & 0.05  \\
20110803 & M6.0       & 133017     & 135858   & 18.7      & 72.5      & -53.8     & 0.81      & 0.98            & 0.81      & 0.98            & 0.78      & 0.99            & 26.7   & 0.25  \\
20110804 & M9.3       & 34215      & 40028    & 6.4       & 4.4       & 2.0       & 1.13      & 0.14            & 1.09      & 0.22            & 1.12      & 0.16            & 5.6    & 0.05  \\
20110808 & M3.5       & 180113     & 181358   & 32.5      & 49.6      & -17.1     & 1.02      & 0.43            & 0.88      & 0.80            & 0.84      & 0.86            & 10.6   & 0.06  \\
20110809 & X6.9       & 75901      & 80629    & -11.4     & -4.0      & -7.4      & 0.68      & 0.96            & 0.65      & 0.98            & 0.69      & 0.96            &        &      
\end{tabular}
\begin{tablenotes}
 \item Note: Table 4 is published in its entirety in a machine readable format. A portion is shown here for guidance regarding its form and content.
\end{tablenotes}
\end{table*}

\bibliographystyle{yahapj}
\bibliography{refs}

\end{document}